\documentclass{elsart}
\usepackage{graphicx}
\usepackage[dvips]{epsfig}
\usepackage{latexsym}
\journal{Digital Signal Processing}
\newcommand{\uc}{{\underline c}}

\newcommand{\R}{I \! \! R}
\newcommand{\N}{I \! \! {N}}

\newcommand{\C}{I \! \! \! \! {C}}

\newcommand{\ute}{{\underline \theta}}

\newcommand{\ut}{{\underline t}}

\newcommand{\ue}{{\underline e}}

\newcommand{\us}{{\underline s}}

\newcommand{\ua}{{\underline a}}

\newcommand{\uga}{{\underline \gamma}}
\newcommand{\ugat}{\tilde{{\underline \gamma}}}
\newcommand{\uxi}{{\underline \xi}}
\newcommand{\uze}{{\underline \zeta}}
\newcommand{\uzet}{\tilde{{\underline \zeta}}}
\newcommand{\ueps}{{\underline \epsilon}}

\newcommand{\argmin}{\mbox{argmin}}

\newcommand{\be }{\begin{eqnarray*}}
\newcommand{\ee}{\end{eqnarray*}}
\newtheorem{propo}{Proposition}

\begin{document}

\begin{frontmatter}

\title{A black box method for solving the complex exponentials approximation problem }
\author{Piero Barone }
\address{Istituto per le Applicazioni del Calcolo ''M. Picone'', C.N.R.\\
via dei Taurini 19, 00185 Rome, Italy\\
e-mail: piero.barone@gmail.com; p.barone@iac.cnr.it}

\begin{abstract}
A common problem, arising in many different applied
contexts, consists in  estimating the number of
exponentially damped sinusoids whose weighted sum best fits a finite set
of noisy data and in estimating their parameters. Many
different methods exist to this purpose. The best of them are based on  approximate Maximum Likelihood
estimators, assuming to know the number of damped sinusoids, which can then be estimated by an order selection procedure.
As the problem can be severely ill posed, a stochastic perturbation method is proposed which provides better results than
Maximum Likelihood based methods when the signal-to-noise ratio is low.
The method depends on some hyperparameters which turn out to be essentially independent of the application.
Therefore they can be fixed once and for all, giving rise to a black box method.

\end{abstract}

\end{frontmatter}

\section*{Introduction}

Let's consider the model  \begin{eqnarray} f_R(t;q,P_R)=
\sum_{j=1}^q A_j\rho_j^t \cos(2 \pi \omega_j t+\theta_j),
\;\;
t\in\R^+,\;\;\omega_h\ne\omega_k\;\forall h,k,\\
P_R=\{A_j,\rho_j,\omega_j,\theta_j,\;j=1\dots,q\}\in\R^{4q}
 \label{modre}\end{eqnarray}
 and assume that we want
to estimate $q,P_R$ from the data
$$a_k=f_R(k\Delta)+\epsilon_k,\;k=0,\dots,n-1, n\ge 4q$$ where $\Delta>0$ is known, $\epsilon_k$ are
i.i.d. zero-mean Gaussian variables with  variance
$\sigma^2$.
In order to make the model $f_R$ identifiable from
$\{a_k\}$ we assume that $|\omega_j|\Delta\le\pi, \;\forall
j$. In fact  if e.g. $\omega_r\Delta>\pi$ there exists
$\tilde{\omega}\in[-\pi,\pi]$ such that
$\omega_r\Delta=\tilde{\omega}\Delta+2\pi h, h\in\N, h\ne
0$ and $f_R(t;q,P_R)=f_R(t;q,P_R')$ where
$P_R'=P_R\setminus \{\omega_r\}\bigcup\{\tilde{\omega}\}$.
We notice that $f_R(t,q,P_R)$ is a particular case of the
complex model \be f(t;p,P)= \sum_{j=1}^p
c_j\xi_j^t,\;\;t\in\R^+,
\\P=\{c_j,\xi_j,\;j=1,\dots,p\}\in\C^{2p}\ee when $p=2q,
q\in\N,\;\Im m(f)=0$ and \be c_j=\frac{1}{2}A_{j}
e^{i\theta_j}, \;\xi_j=\rho_j e^{i 2 \pi\omega_j}, j=1,\dots,q,\\
c_j=\frac{1}{2}A_{j-q} e^{-i\theta_{j-q}},
\;\xi_j=\rho_{j-q} e^{-i 2 \pi\omega_{j-q}}, j=q+1,\dots,p.\ee
Therefore in the following we consider the problem of
estimating $P$ from the complex data
$(a_k,\;k=0,\dots,n-1)$ with the identifiability condition
$|arg(\xi_j)|\Delta\le\pi\;\forall j$, where the noise
$\epsilon_k$ are i.i.d. zero-mean complex Gaussian
variables with  variance $\sigma^2$ i.e. the real and
imaginary parts of $a_k$ are independently distributed as
Gaussian variables with variance $\sigma^2/2$ and mean $\Re
e[f(k\Delta)],\Im m[f(k\Delta)]$ respectively.

The problem described above arises in many fields.
A  not exhaustive list is the
following: noisy Hausdorff moment problem, numerical
inversion of Laplace transform, noisy trigonometric moment problem,
identification of constant coefficients ODE from its transient response, approximation
by complex exponentials functions, modal analysis,
direction of arrival problem, shape from moments problem
\cite{bama1,bama2,barase,emg,hen2,hs0,maba1,scharf}.
The problem belongs to the class of inverse problems. Many references
on the statistical approaches to solve them can be found in \cite{aka}.

In the specific case,
it is well known that the problem can be severely ill posed,
depending on the relative location in the complex plane of the
points $\xi_j,j=1,\dots,p$ and on the ratios
$SNR_j=|c_j|/\sigma,j=1,\dots,p$. A further difficulty is related to
the fact that $p$ is unknown. This means that when the ratios
$SNR_j,j=1,\dots,p$ are bounded by some constant $C<\infty$ even if
you are able to guess the right order $p$ of the model, different
realizations of the process $a_k$ can give rise to quite different
estimates of the other parameters in $P$. The difficulty of guessing
the right order is related to the difficulty of estimating the other
parameters. In fact if these were correctly estimated a good guess
of $p$ would minimize an order selection criterium such as AIC or
BIC \cite{aka}. Unfortunately you cannot hope to get  good estimates of the
other parameters if $p$ is not correctly estimated. Because of this
situation many methods have been proposed to solve the problem by
filtering the noise in different ways and/or considering different
estimators. Those which provide the best performances, assuming to
know the right order $p$, compute an approximation of the Maximum
Likelihood estimator of the parameters filtering somewhat the noise
at the same time \cite{hs0,hs1,hs2}. The guess of the order is then used to build
the noise filter and therefore to improve the estimates of the other
parameters. Different guesses can  be tested in order to
minimize an order selection criterium. A black box procedure can then be devised.

In \cite{dsp10} a
 method which encompasses all
these difficulties was proposed and experimentally compared with standard alternatives on a
few typical problems some of them based on real data. The results were quite good. However the proposed method was not a black box one. Some problem-dependent hyperparameters had to be chosen which made it difficult to appreciate the average quality of the method.
It was noticed in \cite{dsp10}(Remark at pg.4) that one of the most critical hyperparameter is the number of data and some heuristic arguments to justify this fact were provided.
In this paper some theoretical results that support this claim are given. The idea is then to use a statistic related to the stationarity of the residuals to choose among different solutions obtained by using different subsets of the original data set.
For each data subset a  black-box method is proposed, based on a two-steps procedure. The
first step consists of a method to estimate the distribution in the complex plane of the
$\xi_j,j=1,\dots,p$ which are the most critical parameters \cite{distrf}. This allows to identify the subsets of the complex plane which critical parameters are likely to belong to. An important hyperparameter which appear in this step is estimated on the basis of some partially heuristic results.
The second step  makes use of the stochastic perturbation approach given in \cite{j08,dsp10} suitably improved to have better control on the effect of noise. The resulting algorithm still depends on some hyperparameters which however turn out to be  weakly dependent on the specific data set. As a consequence it was possible to fix them once and for all  thus getting a black box method. It was then possible  to perform a simulation study to get information about the average performances of the proposed method for several SNRs. Comparisons were done with one of the best known standard methods (GPOF \cite{hs0}) coupled with BIC for choosing the right order and with the same statistic as above for choosing the best data set. Moreover the method was used to solve two of the problems presented in \cite{dsp10} improving on the results reported there.

The paper is organized as follows. In section 1 the Maximum Likelihood (ML) and related estimators and their properties in this context are shortly reviewed and  the MLE density is studied as a function of the number of data and the noise variance. Moreover a short overview of pencil methods is also given because GPOF algorithm is  used in the proposed method and also for comparisons. In section 2 the proposed method is described and  critical hyperparameters required to make it automatic are discussed and estimated. In section 3  numerical results are reported.

\section{Properties of the Maximum Likelihood Estimator}

\subsection{Algebraic and statistical properties of MLE}
Maximum likelihood estimates $P_{ML}$ of the parameters $P$
of the model $f(t;p,P)$, assuming that $p$ and $\sigma^2$ are known, are
obtained by \be P_{ML}=\mbox{argmax}_{P}\;
e^{-\frac{\|\ua-f(\ut;p,P)\|^2_2}{\sigma^2}}
=\mbox{argmin}_P\;\|\ua-f(\ut;p,P)\|^2_2\ee where
$\ua=[a_0,\dots,a_{n-1}]$,
$\ut=[0,\Delta,\dots,(n-1)\Delta]$. In order to solve this
nonlinear least squares problem, following \cite{gp}, we notice
that the problem is separable. In fact we can split the
parameters $P$ in two sets $P=P_c\bigcup P_\xi$ where $
f(t;p,\uga,\uze)= \sum_{j=1}^p \gamma_j\zeta_j^t$. For each
fixed value $\uze\in P_\xi$ let us consider the function
$\uga(\uze)$ defined by \be
\uga(\uze)&=&\mbox{argmin}_\uga\|\ua-f(\ut;p,\uga,\uze)\|^2_2=
\mbox{argmin}_\uga(\ua-V\uga)^H(\ua-V\uga)\\&=&
(V^HV)^{-1}V^H\ua\ee
where $V=V(\uze)$ is the Vandermonde matrix  of order
$n\times p$ of the vector $\uze$, $H$ denotes transposition
plus conjugation and $I_n$ is the identity matrix of order
$n$. It is proved in \cite{gp} that, substituting $\uga(\uze)$
in $\|\ua-f(\ut;p,\uga,\uze)\|^2_2$ and minimizing w.r.to
$\uze$, we get \be \uxi_{ML}=\mbox{argmin}_\uze
\|\ua-f(\ut;p,\uga(\uze),\uze)\|^2_2= \\
\mbox{argmin}_\uze
(\ua-V(V^HV)^{-1}V^H\ua)^H(\ua-V(V^HV)^{-1}V^H\ua) =\\
\mbox{argmin}_\uze\ua^H(I_n-V(V^HV)^{-1}V^H)\ua \ee and \be
\uc_{ML}=\uga(\uxi_{ML}).\ee In order to study the
properties of the ML estimator we start by noticing that
\begin{propo}
It does not exist an efficient estimator of the parameters
$P$. Specifically the MLE of $P$ is not efficient.
\label{pro1}
\end{propo}
\noindent \underline{Proof.} We notice that the
log-likelihood function is an absolutely  continuous function
of $P$. Hence, by Corollary 3.1 and Theorem 3.1 of \cite{jos}
 if the variance of an estimator of $P$ would
attain the Cramer-Rao bound this would imply that the
probability density
$$\frac{1}{(\pi \sigma^2)^n} e^{-\frac{\|\ua-f(\ut;p,P)\|^2_2}{\sigma^2}}
$$ of $\ua$ would belong to the exponential family. But
this is false because of the dependence of $f(t;p,P)$ on
$\xi_j^t,\;j=1,\dots,p$ which make it impossible to factorize the argument of the exponential
in the product of two functions which depend only on
the parameters and the observation variable respectively. $\Box$

\subsection{Approximate MLE: complex exponentials interpolation}
 We then consider the problem of
interpolating the data $\ua$ by
 means of a linear combination of complex exponential
 functions $\tilde{\zeta}_j^t,\;\;\tilde{\zeta}_j\in\C,
 \;\;j=1,\dots,n/2$,  that is
 to find $n$ complex numbers
 $[\ugat,\uzet]=\{\tilde{\gamma}_j,\tilde{\zeta}_j\},j=1,\dots,n/2$
 such that $\ua=V(\uzet)\ugat$. In the following the complex exponentials interpolation problem will be denoted by CEIP. Equivalently (see e.g. \cite{hen2,bama2}) we could
 consider the problem of building the Pade' approximation $[n/2,n/2-1]$ to the
 $Z-$transform of $a_k,k=0,1,\dots$.
To this aim let us consider the Hankel matrix pencil $
U_1-zU_0,\;\;z\in\C $ where
$$U_0(\underline a)=U(a_0,\dots,a_{n-2}),\;\;\;\;
  U_1(\underline a)=U(a_1,\dots,a_{n-1})$$ and
\be U(x_1,\dots,x_{n-1})=\left[\begin{array}{llll}
x_1 & x_{2} &\dots &x_{n/2} \\
x_{2} & x_{3} &\dots &x_{n/2+1} \\
. & . &\dots &. \\
x_{n/2} & x_{n/2+1} &\dots &x_{n-1}
  \end{array}\right]\ee
It is well known (e.g.\cite{hen2}) that, provided that $\det U_0\ne 0,
\det U_1\ne 0$, a unique solution of CEIP exists which is given by
$\uzet=\uxi_{GE}$, where $\uxi_{GE}$ are the generalized
eigenvalues of the pencil $U_1-zU_0$ and
$\ugat=W_{GE}^T\ua$ where $W_{GE}$ is the matrix of
generalized eigenvectors of $U_1-zU_0$ and $T$ denotes
transposition. Moreover it turns out that
$W_{GE}=\tilde{V}(\uxi_{GE})^{-T}$ where
$\tilde{V}(\uxi_{GE})$ is the square Vandermonde matrix
based on $\uxi_{GE}$. These properties can be easily
checked by noticing that if $\ua=V(\uzet)\ugat$ then \be
U_0=\tilde{V}(\uzet)C\tilde{V}(\uzet)^T,\;\;
U_1=\tilde{V}(\uzet)CZ\tilde{V}(\uzet)^T\ee where \be
C=\mbox{diag}\{\tilde{\gamma}_1,\dots,\tilde{\gamma}_{n/2}\}
\mbox{ and }
Z=\mbox{diag}\{\tilde{\zeta}_1,\dots,\tilde{\zeta}_{n/2}\}\ee and
therefore $U_1 \tilde{V}(\uzet)^{-T}=U_0
\tilde{V}(\uzet)^{-T} Z$ which implies that $\uzet$ are the
generalized eigenvalues of the pencil $U_1-zU_0$.
The relation between $[\uc_{ML},\uxi_{ML}]$ and
$[\ugat,\uzet]$ is given by
\begin{propo}
If $n=2p$ then $[\uc_{ML},\uxi_{ML}]= [\ugat,\uzet].$
\label{pro1}
\end{propo}
\noindent \underline{Proof.} Let be $V=V(\uzet)$.
Substituting $\ua=V\ugat$ in $
\ua^H(I_n-V(V^HV)^{-1}V^H)\ua$ we get \be \ugat^H
V^H(I_n-V(V^HV)^{-1} V^H)V\ugat=0.\ee But $
\ua^H(I_n-V(V^HV)^{-1}V^H)\ua\ge 0,$ hence
$\|\ua-f(\ut;p,\uga(\uze),\uze)\|^2_2$ takes its least
possible value when $V=V(\uzet)$ therefore
$\uzet=\uxi_{ML}$ and
$\ugat=(V^HV)^{-1}V^H\ua=\uc_{ML}.\;\;\Box$

\subsection{Bias of MLE}
We show now
that the MLE can not have moments. In particular MLE can not
have the mean, therefore bias can not be defined. Let us
consider the case when
$n=2,p=1,\theta_1=\omega_1=0,|\rho|<1$. Therefore
$$a_0=A+\epsilon_0,\;a_1=A\rho+\epsilon_1,\;\;U_0=a_0,\;U_1
=a_1,\;\;\rho_{ML}=\frac{a_1}{a_0}.$$ The density of
$\rho_{ML}$ is then the density of the ratio of two
independent Normal variables with means $A$ and $A\rho$
respectively and  variance $\sigma^2$ which is given by

$$p_2(x)=\frac{1}{2\pi \sigma^2} \int_{-\infty}^\infty\int_{-\infty}^\infty \delta\left(x-\frac{w}{v}\right) e^{-\frac{(v-A)^2+(w-A\rho)^2}{2\sigma^2}} dvdw.$$
We notice that $\rho_{ML}$ is also the generalized eigenvalue of the pencil $a_1-za_0$ and  $p_2(x)$ can be rewritten as
$$p_2(x)=E[\delta(x-\rho_{ML})],$$
therefore $p_2(x)$ is the condensed density of the generalized eigenvalue $\rho_{ML}$ (see e.g. \cite{j05}).
By performing the change of variables
$$\lambda=v,\;\;\;\mu=\frac{w}{v}$$ we get
$$p_2(x)=\frac{e^{-\frac{A^2 (\rho-x)^2}{2 \sigma^2 \left(x^2+1\right)}} A( \rho x+1)
   \mbox{Erf}\left(\frac{ A( \rho x+1)}{\sqrt{2\sigma^2\left(x^2+1\right)}} \right)}{\sqrt{2 \pi \sigma^2}
   \left(x^2+1\right)^{3/2}}+\frac{e^{-\frac{A^2 \left(\rho^2+1\right)}{2 \sigma^2}}}{\pi  \left(x^2+1\right)}$$
which is a Cauchy-like density and therefore moments do not
exist. We can expect that the same problem arises for the general model. We cannot therefore define the bias. However we can define instead the quantity $\|M[\uxi_{ML}]-\uxi\|$ where $M[\uxi_{ML}]$ is the principal mode of the condensed density of the generalized eigenvalues and for simplicity we call it "bias" of $\uxi_{ML}$. We have
\begin{propo}
When $n=2$,  $A_{ML},\rho_{ML}$ are biased.
\end{propo}
\noindent \underline{Proof.}
 Let us assume that $A>0,\;\rho>0$. By \cite{pham}[Prop.7] for $x>0$, $p_2(x)$ has a unique mode not greater than $\rho$. Moreover  we notice that
$$\lim_{\sigma^2\rightarrow 0}p_2(x)=\delta(x-\rho)$$
and
$$\lim_{\sigma^2\rightarrow \infty}p_2(x)=\frac{1}{\pi \left(x^2+1\right)}.$$
Therefore  the mode of $p_2(x)$  moves continuously from $\rho$ to $0$ as $\sigma^2$ moves from $0$ to $\infty$. Hence
the bias of $\rho_{ML}$ is zero only when $\sigma=0$ and, as $A_{ML}=\frac{a_0+\rho_{ML}a_1}{1+\rho_{ML}^2}$, the same is true for $A_{ML}.$
By using a similar argument the same result can be proved also when the hypothesis  $A>0,\;\rho>0$ is relaxed$.\;\;\;\Box$

We can argue that the same kind of conclusion holds in the general case. Moreover
 one
could argue that  when $\sigma>0$,  for $p$ fixed, the bias is a decreasing function of $n$.
 This is not
the case as we now show for the simplest case of the model
$a_t=A\rho^{(t-1)}+\epsilon_t,\;\;t=0,\dots,n-1$ where
$|\rho|<1$ and $\epsilon_t$ are i.i.d. Gaussian zero-mean
random variables with variance $\sigma^2$. We notice that the case $|\rho|<1$ is the critical one because when $|\rho|>1$ the MLE of $\rho$ are trivially asymptotically unbiased as the noise will become negligible w.r. to the signal if $n=n(\sigma)$ is large enough.  The following Proposition holds, where for simplicity the approximated density is denoted as the true one.

\begin{propo}
When $|\rho|<1$ the density of the MLE of  $\rho$ can be
approximated by a density $p_n(x)$ such that
$$\lim_{n\rightarrow\infty}p_n(x)=0 \mbox{ if }|x|\ge 1$$
$$\lim_{n\rightarrow\infty}p_n(x)=p_\infty(x),
\;x\in(-1,1)$$ where $p_\infty(x)$ is a density such that
$$\lim_{\sigma\rightarrow 0}p_\infty(x;\rho,\sigma)=\delta(x-\rho)$$
(in the sense of distributions). For $\sigma>0$,
$p_\infty(x)$ has at most one mode in $(-1,1)$ and two vertical asymptotes in $\pm 1$.
Moreover, for $\sigma>0$, $p_n(x)$ has two relative maxima whose values tend to infinity as
$n\rightarrow\infty.$
\end{propo}

\noindent \underline{Proof.} Let us denote by $\ute_{ML}=[\rho_{ML},A_{ML}]$
the ML estimates of $\rho$ and $A$ and by $\ute^*=[\rho,A]$ the true parameters. With the notations used before, the model can be written in vector form as
$$ \ua=\us(\ute^*)+\ueps,\;\;\ueps\sim N(\underline{0},\sigma^2 I_n)$$
where $s(t;\ute^*)=A\rho^{(t-1)}.$
The log-likelihood function  is
$$L(\ute)=-\frac{1}{2 \sigma^2}(\ua-\us(\ute))^T(\ua-\us(\ute))$$
and the ML estimates $\ute_{ML}$ satisfy the nonlinear system
$$G(\ute_{ML})\{\ua-\us(\ute_{ML})\}=\underline{0},\;\;\;G_{hk}(\ute)=\frac{\partial}{\partial\ute_h} s_k(\ute),\;\;h,k=1,\dots,n.$$
Following \cite{acb}, if we  consider the  first order
Taylor series approximation of $\us(\ute)$ around the point  $\ute_e$ where we want to approximate the density, we get $$\us(\ute)\approx\us(\ute_e)+G(\ute_e)^T[\ute-\ute_e]$$ and the nonlinear system becomes the linear one
$$G(\ute_e)\{\ua-\us(\ute_e)-G(\ute_e)^T[\ute-\ute_e]\}=G(\ute_e)[\ua-\us(\ute_e)]-
G(\ute_e)G(\ute_e)^T[\ute-\ute_e]=\underline{0}$$
whose solution is
$$\tilde{\ute}_{ML}=\ute_e+[G(\ute_e)G(\ute_e)^T]^{-1}G(\ute_e)[\ua-\us(\ute_e)]$$
which is a linear function of the Gaussian data $\ua$ and therefore
$\ute_{ML}$ is approximately distributed as a Gaussian vector with mean $$\mu_\theta=[G(\ute_e)G(\ute_e)^T]^{-1}G(\ute_e)[\us(\ute^*)-\us(\ute_e)]$$ and covariance
$$\Sigma_\theta=\sigma^2G(\ute_e)^T[G(\ute_e)G(\ute_e)^T]^{-1}G(\ute_e).$$ Hence $\ute_{ML}\sim N(\mu_\theta,\Sigma_\theta)=p_n(\ute_e)=p_n(\rho_e,A_e)=p_n(x,y).$
We notice that the parameter $y$ can be factored out in $p_n(x,y)$. In fact
$$p_n(x,y)=y a_n(x) e^{y^2 b_n(x)+y c_n(x) -d_n(x)}$$  where, dropping the dependence on $x$ for simplicity
\begin{eqnarray*}\lefteqn{a_n=\frac{1}{2 \pi  \sigma^2}\sqrt{\delta_3},\;\;b_n=\frac{\delta_6}{2 \sigma^2 },\;\;c_n=\frac{A \delta_2}{\sigma^2 },\;\;d_n=\frac{A^2}{2 \sigma^2 \delta_3} \left[\delta_2 (\delta_1 \delta_2 - \rho \delta_4 \delta_5) - \rho \delta_5 (\delta_2 \delta_4 - \rho \delta_5 \delta_6)\right]}\end{eqnarray*}
and $\delta_j,\;j=1,\dots,6$ are polynomials in $x$:
\begin{eqnarray*}\lefteqn{\delta_1=\sum_{j=0}^{n-2}(j+1)^2 x^{2j}=\frac{x^{2( n-1)} \left(\left(-n
   x^2+n+x^2\right)^2+x^2\right)-
   \left(x^2+1\right)}{
   \left(x^2-1\right)^3}}
\end{eqnarray*}
\begin{eqnarray*}\lefteqn{\delta_2=\sum_{j=0}^{n-1} (\rho x)^{j}=\frac{(\rho x)^{n}-1}{\rho x-1}}
\end{eqnarray*}
\begin{eqnarray*}\lefteqn{\delta_3= {n+1 \choose 3}x^{2(n-2)}+\sum_{j=0}^{n-3}{3+j \choose 3}(x^{2j}+x^{2(2 n-4-j)})=\frac{-\left(n^2 \left(x^2-1\right)^2+2 x^2\right)
   x^{2 n}+x^{4 n+2}+x^2}{x^2 \left(x^2-1\right)^4}}
\end{eqnarray*}
\begin{eqnarray*}\lefteqn{\delta_4=\sum_{j=0}^{n-2}(j+1)x^{2j+1}=\frac{\left((n-2) (x^2-1)+x^2-2\right) x^{2
   (n-2)+3}+x}{\left(x^2-1\right)^2}}
\end{eqnarray*}
\begin{eqnarray*}\lefteqn{\delta_5=\sum_{j=0}^{n-2} (j+1)(\rho x)^{j}=\frac{ \left((n-1) \rho x-n-4\right) (\rho x)^{n-1}+1}{(\rho x-1)^2},\;\;\delta_6=\sum_{j=0}^{n-1} x^{2 j}=\frac{x^{2 n}-1}{x^2-1}}
\end{eqnarray*}
Therefore the approximated density of $\rho_{ML}$ is given by
$$p_n(x)=\int_{-\infty}^\infty p_n(x,y)dy=
 \frac{\sqrt{\pi} a_n c_n}{2 \sqrt{b_n^3}} e^{\frac{c_n^2}{4b_n} - d_n}.$$
We notice that $p_n(x;\rho)=p_n(-x;-\rho)$ because this property holds for $\delta_2$ and $\delta_5$ as they are functions of $\rho x$.
For $n\rightarrow \infty$ we get
\begin{eqnarray*}\lefteqn{a_\infty=\frac{1}{2 \pi  \sigma^2
   \left(x^2-1\right)^2},\;\;b_\infty=\frac{1}{2 \sigma^2
   \left(1-x^2\right)},\;\;c_\infty=\frac{A}{\sigma^2 (1-\rho x)}}
\end{eqnarray*}
\begin{eqnarray*}\lefteqn{d_\infty=-\frac{A^2 \left(x^2-1\right)
   \left(\rho^2
   \left(x^2+1\right)-4 \rho
   x+x^2+1\right)}{2 \sigma^2 (\rho
   x-1)^4}}
\end{eqnarray*}
and hence
\begin{eqnarray*}p_\infty(x)=
\frac{\sqrt{\pi} a_\infty c_\infty}{2 \sqrt{b_\infty^3}} e^{\frac{c_\infty^2}{4b_\infty} - d_\infty}=
\frac{A \exp \left(-\frac{A^2
   \left(1-x^2\right)
   (\rho-x)^2}{2 \sigma^2 (\rho
   x-1)^4}\right)}{\sqrt{2 \pi
   } \sigma \sqrt{1-x^2} (1-\rho x)}
\end{eqnarray*}
We notice that $p_\infty(x)$ assumes real values only for $x\in(-1,1)$ and has two poles in $\pm 1$.
 We also have
$$\lim_{\sigma\rightarrow 0}p_\infty(x)=\left\{
\begin{array}{ll}
0\;\;\;\; x\ne \rho,\;x\in(-1,1) \\
\infty\;\;\;\;x=\rho \end{array} \right.$$
   therefore
$$\lim_{\sigma\rightarrow
0}p_\infty(x;\rho,\sigma)=\delta(x-\rho)$$ in the weak
sense.

\noindent By taking the logarithm we get
$$\log[p_\infty(x)]=-\frac{A^2 \left(1-x^2\right) (\rho-x)^2}{2 \sigma^2 (\rho
   x-1)^4}+\log (A)-\log (1-\rho x)-\frac{1}{2} \log
   \left(2 \pi  \sigma^2\right)-\frac{1}{2} \log
   \left(1-x^2\right)$$
whose derivative is the rational function
$$\frac{A^2 \left(x^2-1\right) (\rho-x) \left[\rho^2
   \left(x^2-2\right)+2 \rho x-2 x^2+1\right]+\sigma^2
   \left[\rho \left(2 x^2-1\right)-x\right] (\rho
   x-1)^4}{\sigma^2 \left(1-x^2\right) (\rho x-1)^5}.
$$
with poles  $\pm 1$ and  $1/\rho$ not in $(-1,1)$. The numerator can be decomposed as
$$L_1(x)+\sigma^2 L_2(x)$$
with $L_2'(x)=(\rho x-1)^3 \left[4 \rho^2 \left(3 x^2-1\right)-9 \rho x+1\right]$, hence $L_2(x)$ has only one stationary point  in $(-1,1)$.
The roots of $L_1(x)$ are
\begin{eqnarray*}\lefteqn{x_1=\rho,\;\;\;x_{2,3}=\pm 1,\;\;\;x_4=\frac{\sqrt{2}
   \left(\rho^2-1\right)+\rho}{2-\rho^2},\;\;\;x_5=\frac{\sqrt{2}
   \left(\rho^2-1\right)-\rho}{\rho^2-2}}\end{eqnarray*}
 where $L_1'(x_1)<0$ (relative maximum of $p_\infty(x)$) and $L_1'(x_{4,5})>0$ (relative minima of $p_\infty(x)$). As all the roots are real the zeros of $L_1(x)$ interlaces with those of $L_1'(x)$ (see e.g. \cite{fisk}) therefore between two consecutive zeros $L_1(x)$ can't have more than one relative maximum or minimum. Hence as  the numerator of $\frac{d\log[p_\infty(x)]}{dx}$ is obtained from $L_1(x)$ by adding the  perturbation $\sigma^2 L_2(x)$,  the number of its zeros is not greater than three in $(-1,1)$ and by continuity their value is as close to
   $x_1,x_4,x_5$ as $\sigma^2$ is small. Summing up $p_\infty(x)$ has two vertical asymptotes in $\pm 1$ and at most one relative maximum as close to $\rho$ as $\sigma^2$ is close to zero. We can therefore expect that if $\sigma>0$, for $n\rightarrow \infty$ the density of $p_n(x)$ is concentrated close to $\pm 1$. In fact it is easy to see that also the first derivative of $\log[p_n(x)]$ is a rational function $\frac{P_n(x)}{Q_n(x)}$ and
$$Q_n(x)=\sigma^2 \delta_2 \delta_6^2 \delta_3^2$$
has no roots in the interval $(-1/\rho,1/\rho)\supset(-1,1)$. Therefore the vertical asymptotes of $p_\infty(x)$ in $\pm 1$ can not  be approximated by asymptotes of $p_n(x)$ i.e. by zeros of $Q_n(x)$.  Therefore we can expect that $P_n(x)$ has two zeros which approach $\pm 1$ as $n\rightarrow\infty$ and which correspond to two local maxima whose values tend to infinity. $\Box$

We notice that, because the sampling interval $\Delta$ is fixed, increasing $n$ has the same effect of increasing the noise variance $\sigma^2$. Moreover, after Proposition 2, when $n=2p$ the MLE density is equal to the condensed  density of the generalized eigenvalues of the pencil $U_1-zU_0$. The behavior of this function was studied in \cite{j08} as a function of $\sigma$ in the general case of complex exponential functions perturbed by Gaussian complex noise. When $\sigma\downarrow 0$ the condensed density tends weakly to a sum of $p$ Dirac's delta  centered on the true values $\xi_j,j=1,\dots,p$. When the signal is zero or, equivalently, when $\sigma\downarrow \infty$ the condensed density is such that in polar coordinates the phase is uniformly distributed in $[-\pi,\pi]$ and the modulus has a Dirac delta distribution centered on $1$. Moreover in \cite{pde} it was shown that the condensed  density of the generalized eigenvalues satisfies a parabolic partial differential equation where $\sigma$ plays the role of time. This PDE then rules the diffusion of the condensed density from the sum of $p$ Dirac's delta centered on the true values to the Dirac's 2d-measure centered on the unit circle.

\begin{figure}
\begin{center}
\hspace{1.7cm}{\fbox{\psfig{file=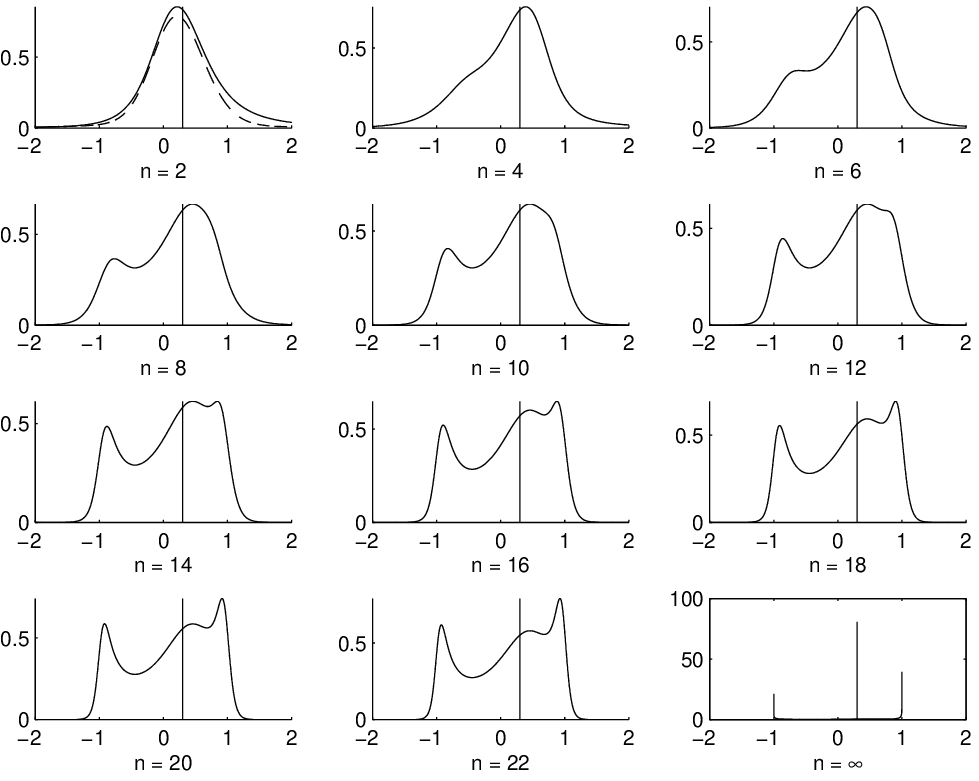,height=8cm,width=16cm}}}
\end{center}
\caption{ \label{lfig00}}   Approximation of the  density of the MLE of $\rho$ as a function of the number of data when  $\sigma=0.5,\;\rho=0.3,\;A=1$. For $n=2$ the true density is also plotted (dashed). The true value of $\rho$ is represented by the vertical bar.
\end{figure}

\begin{figure}
\begin{center}
\hspace{1.7cm}{\fbox{\psfig{file=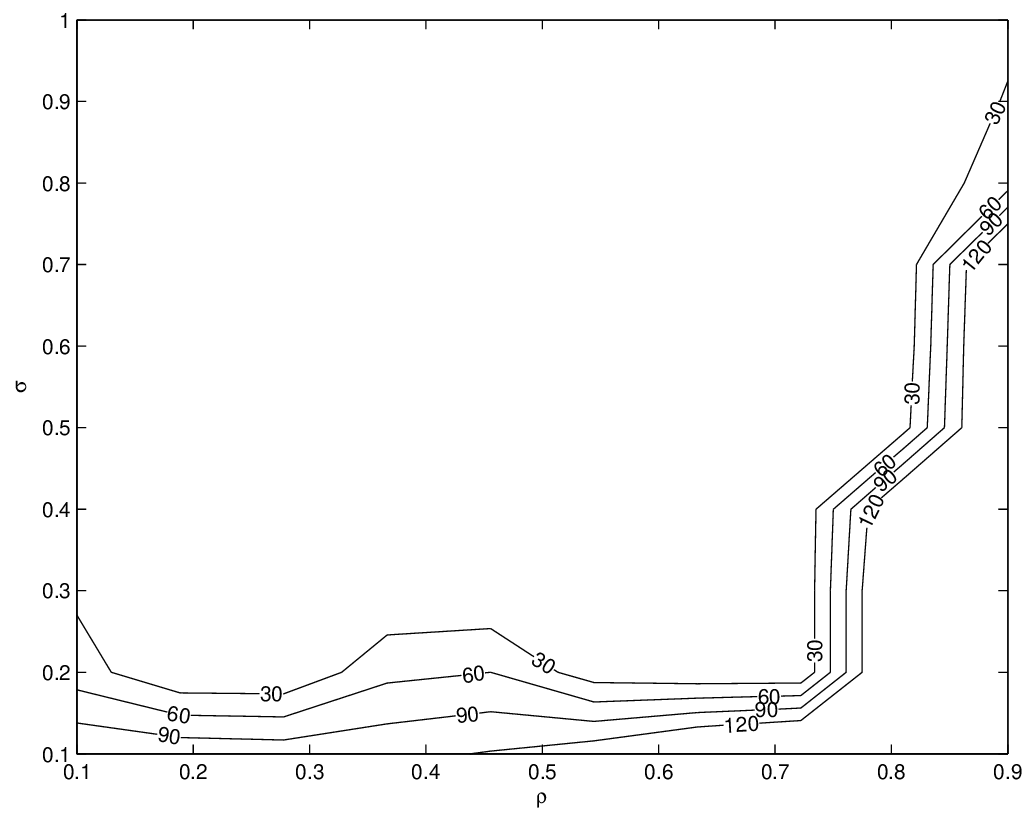,height=8cm,width=16cm}}}
\end{center}
\caption{ \label{lfig0}} Optimal value of $n$ as a function of $\rho$ and $\sigma$ when $A=1$.
\end{figure}

Therefore the claims of Proposition 4, even if obtained through an approximation, are consistent with these general results. In fig. \ref{lfig00} the behavior of $p_n(x)$  is shown.
We see also that $p_n(x)$ is unimodal  for small values of $n$, and the mode is close to the true value of $\rho$. By increasing $n$ two secondary modes appear close to $\pm 1$, and for $n$ large enough one of the secondary modes becomes the principal one.  We conclude that for each $\sigma, A, \rho$ there exists an optimal value of $n$ which minimizes $|\rho-\hat{\rho}(n;\sigma, A, \rho)|$. This value was computed numerically by using the approximated density $p_n(x)$ of the ML estimator of $\rho$ for  $A=1$ and for several values of $\rho>0$ and $\sigma$  and plotted in fig. \ref{lfig0}. Only positive values of $\rho$ are considered because $p_n(x;\rho)=p_n(-x;-\rho).$ As expected, the optimal value of $n$ is an increasing function of $|\rho|$ and a decreasing function of $\sigma$. We can conjecture that this conclusion holds for the general model too. The choice of the number of data is critical: not less than $2p$ data must be used but, unfortunately, there is not an easy way to estimate the optimal value of $n$ also because it depends on the true unknown parameters. In the following we propose an estimation procedure where this problem is taken into account.

\subsection{Standard pencil methods: GPOF}
Computation of MLE is usually complicated because the right
model order $p$ should be known and many local maxima are
present when SNR is low or moderately large.
 In literature many algorithms to get approximate MLE exist
  and their relative merits are usually measured in
terms of the CR bound for the asymptotic unbiased
estimators \cite{bremac,kay}. This does not make much sense
because we are interested in solving the problem when
$\sigma>0$ but can help to compare algorithms. As expected
because of the asymptotic unbiasdness, when the noise
variance is less than a threshold, all algorithms produce
reasonable estimates (see \cite{emg} for a comparison). Moreover
some heuristic algorithms can exceed the CR bound (because
of the bias) and hence it is suggested that the bias can
help to decrease the noise threshold below which meaningful
estimates can eventually be computed \cite{kay}.
Because of the connection between ML estimation
and complex exponential interpolation, many approximate ML
algorithms are based on complex exponential interpolation
of the data. The main advantages over the exact MLE
algorithms are that no initialization must be provided and
the computation is faster. The best of them include some
sort of noise filtering in order to increase the SNR ratio.
Cadzow method \cite{cad} and GPOF \cite{hs0} are examples of this approach.
We give here a short summary of GPOF method because it is used in the proposed estimation procedure described in Section 2 and it will
be used  for comparisons  in Section 3.

Assuming that the
data $\ua$ are noisy and that we know the true number $p$
of complex exponentials, the aim of GPOF is to estimate the
non linear parameters $\xi_j,\;j=1,\dots,p$ by solving a
filtered generalized eigenvalue problem. When the data are
noiseless we know that we can retrieve  $\uxi$ by solving
the CEIP based on a
square pencil of order $p\times p$ i.e. $n=2p$ data are
enough. If we use $n>2p$ data and use a square pencil of
order $n/2\times n/2$  the conditions $\det U_0\ne 0,
\det U_1\ne 0$  to solve the problem and to get a unique
solution are no longer satisfied because
$\mbox{rank}(U_0)=\mbox{rank}(U_1)=p<n/2$.  When noise is present it makes
sense to assume that $n/2-p$ terms of the model represent
the noise. Therefore we can solve the CEIP of order $n/2$ and then discard the
$n/2-p$ terms associated e.g. with the lowest absolute
values $|c_j|$ of the weights. As an alternative we can
first filter-out the noise from the pencil and then solve a
CEIP of order $p$.
More generally we can assume that the model is made up of
$l$ terms, $l-p$ of them representing the noise, with $p\le
l \le n-p$, i.e. $\ua=V(\uzet)\ugat$ where
$V(\uzet)\in\C^{(n-l)\times l}$ is the Vandermonde matrix
based on $\uzet_j,\;j=1,\dots,l$. We notice that the larger
$l$ the smaller the number of equations $n-l$ that we can
form with $n$ observations. By choosing $l$ we can control
how accurately to represent the noise and hence the signal,
but the price to pay is on the number of constraints that can be considered.
Therefore, when $p$ is unknown, GPOF depends on two hyperparameters $(l,\tilde{p})$ with $\tilde{p}\le l \le n-\tilde{p}$ where  $\tilde{p}$ is a guess of $p$.

  We can then consider a rectangular
pencil $U_1-zU_0$ with \be
U_0=\tilde{V}_1(\uzet)C\tilde{V}_2(\uzet)^T,\;\;
U_1=\tilde{V}_1(\uzet)CZ\tilde{V}_2(\uzet)^T\ee where
$\tilde{V}_1(\uzet)\in\C^{(n-l)\times l},
\tilde{V}_2(\uzet)\in\C^{l\times l}$ are the Vandermonde
matrices based on $\uzet_j,j=1,\dots,l$ and
 \be
C=\mbox{diag}\{\tilde{\gamma}_1,\dots,\tilde{\gamma}_{l}\} \mbox{
and } Z=\mbox{diag}\{\tilde{\zeta}_1,\dots,\tilde{\zeta}_{l}\}\ee
and therefore $U_1 \tilde{V}_2(\uzet)^{\ddagger}=U_0
\tilde{V}_2(\uzet)^{\ddagger} Z$ where
$X^\ddagger=(X^\dagger)^T=(X^T)^\dagger$ and $X^\dagger$
denotes the generalized inverse of $X$. Therefore $\uzet$
are the generalized eigenvalues of the rectangular pencil
$U_1-zU_0$. We want now to compute the signal related
generalized eigenvalues by solving an eigenvalue problem of
order $\tilde{p}$. To this aim let us define the data matrix
 \begin{eqnarray}
U=\left[\begin{array}{llll}
a_0 & a_{1} &\dots &a_{l} \\
a_{1} & a_{2} &\dots &a_{l+1} \\
. & . &\dots &. \\
a_{n-l-1} & a_{n-l} &\dots &a_{n-1}
  \end{array}\right]\in \C^{(n-l) \times (l+1)},\;\tilde{p}\le l \le
  n-\tilde{p}, \label{matu}
\end{eqnarray} from which we can retrieve $U_0, U_1$ by
\begin{eqnarray}U_0=UE_0,\;U_1=UE_1,\;E_0=[\ue_1,\dots,\ue_l],\;
E_1=[\ue_2,\dots,\ue_{l+1}] \label{e0e1}
\end{eqnarray}
 where $\ue_j$ is the $j-$th
column of the identity matrix $I_{l+1}$. Let us consider
then its singular value decomposition $U=PDQ,\;P\in
\C^{(n-l) \times (n-l)},\;D\in \C^{(n-l) \times
(l+1)},\;Q\in \C^{(l+1) \times (l+1)}$. In the noiseless
case $\mbox{rank}(U)=p$ therefore the last $n-l-p$ elements on the
diagonal of $D$ are zero and $U=P^FD^FQ^F$ where $D^F\in
\C^{p \times p}$ is obtained  from $D$ by dropping the last
$n-l-p$ rows or columns,  $P^F\in \C^{(n-l) \times p}$ is
obtained from $P$ by dropping the last $n-l-p$ columns and
$Q^F\in \C^{p \times (l+1)}$ is obtained from $Q$  by
dropping the last $n-l-p$ rows. In the noisy case we can
filter out the smallest $n-l-\tilde{p}$ elements on the diagonal of
$D$ setting them to zero. But then the Hankel structure of
$U^F=P^FD^FQ^F$ is lost. Cadzow \cite{cad} suggests to retrieve
this structure while filtering out the smallest singular
values by the iteration:
\begin{itemize}
\item $U^{(0)}=U$
\item $\mbox{ for }k=0,1,\dots $
\item $\;\;\;\;\;\;\;\;U^{(k)}=P^{(k)}D^{(k)}Q^{(k)}$
\item $\;\;\;\;\;\;\;\;U^F=(P^{(k)})^F (D^{(k)})^F (Q^{(k)})^F$
\item $\;\;\;\;\;\;\;\;U^{(k+1)}=\mbox{ Hankel(}U^F)$
\item $\;\;\;\;\;\;\;\;\mbox{ if } \|U^{(k+1)}-U^{(k)}\| < \eta\;\mbox{then
stop}$
\item end
\end{itemize}
where $\eta>0$ is a small tolerance and the operator Hankel($A$)
maps the matrix $A$ into the matrix obtained by substituting each
element of a secondary diagonal of $A$ by the average of the
elements of that diagonal. In \cite{cad} is proved that this iteration is a
specific instance of a general method which converges under
hypotheses that are verified in the case considered here.
We notice that the iteration can be seen as a filtering algorithm for the data $a_0,\dots,a_{n-1}$ which form the first row and the last column of $U$, the filtered data after $k$ steps being the first row and last column of $U^{(k+1)}.$

Denoting
by $P^FD^FQ^F$ the singular value decomposition of the Hankel matrix
produced by the iteration we have to solve the
rectangular $(n-l)\times l$ generalized eigenvalue problem
$$P^FD^FQ^FE_1W=P^FD^FQ^FE_0WZ.$$
We notice  that $\tilde{P}=P^FD^F\in \C^{(n-l) \times \tilde{p}}$ has
maximum rank $\tilde{p}$ therefore its generalized inverse is
$\tilde{P}^\dagger=(\tilde{P}^H\tilde{P})^{-1}\tilde{P}^H.$
Therefore by left-multiplying by $\tilde{P}^\dagger$ the problem
above reduces to the rectangular $\tilde{p} \times l$ generalized eigenvalue
problem
\begin{eqnarray} Q^FE_1W=Q^FE_0WZ \label{peq}\end{eqnarray}
whose solution is given
by the non-zero eigenvalues of $\tilde{Q}_0^\dagger\tilde{Q}_1\in \C^{l \times l}$
where  $\tilde{Q}_0=Q^FE_0\in \C^{\tilde{p} \times
l},\;\;\tilde{Q}_1=Q^FE_1\in \C^{\tilde{p} \times l}.$ By exploiting the property that the non-zero eigenvalues of $AB$ and $BA$ are the same if
$A\in \C^{m \times n}$ and $B\in \C^{n \times m}$, the signal related generalized eigenvalues of equation (\ref{peq}) are the eigenvalues of
$\tilde{Q}_1\tilde{Q}_0^\dagger\in \C^{\tilde{p} \times \tilde{p}}.$

We notice that this solution slightly differs from the standard one where the singular value decomposition of $U_0$ instead of that of $U$ is considered.
The generalized eigenvalue problem to solve is then
$$U_1^FW=P^FD^FQ^FWZ$$
whose solution is provided by the non-zero eigenvalues of $$(P^FD^FQ^F)^\dagger U_1^F=(Q^F)^H(D^F)^\dagger (P^F)^H U_1^F\in \C^{l \times l}$$ or by the eigenvalues of
$$(D^F)^\dagger (P^F)^H U_1^F (Q^F)^H\in \C^{\tilde{p} \times \tilde{p}}.$$
The solution of equation (\ref{peq}) provides slightly better results only  when the SNR is low and the  improvement is too small to modify the conclusions of a simulation. Therefore
in the following the standard formula is used because it is more convenient from the computational point of view as it does not require the computation of the generalized inverse of a full matrix.

We notice also that the singular value decomposition of $U$ can be
replaced by its $PRQ$ rank revealing decomposition \cite{ch} where $P$
and $Q$ are unitary matrices and $R$ is a trapezoidal matrix such
that the absolute values on the diagonal are in decreasing order. In
fact it turns out that in the noiseless case $R$ is a trapezoidal
matrix of rank $p$ \cite[Section 7.3]{horn} and
noise filtering can be performed by setting to zero the last $n-l-\tilde{p}$
rows of $R$. Despite the obvious computational advantages this
method is worse than the one based on svd for low SNRs because the
best approximation property of svd does not hold.

\section{The proposed method}

\subsection{Outline}
From the discussion of the previous section, in order to
propose a black box method which improves on the bias
affecting exact and approximate MLE, we start from the
CEIP, which is likely
to capture the best features of MLE and exploits the
ensemble behavior (as specified below) of its solution
which is easier to study than the ensemble behavior of
MLE. Specifically the basic observation which motivates the
proposed method is the following. When SNRs are moderate or
low the performances of a good standard algorithm, such as
e.g. GPOF, measured by the MSE of the parameters vary
significantly as a function of the noise realization used.
For example for some noise realizations, two modes with close
frequencies can be well separated even if SNRs are low,
while for other noise realizations, with the same variance,
this is not true. This means that the bias of the frequency
estimates in some cases makes the two modes even closer than
they are making it impossible to separate them while in
other cases the opposite is true. The idea is then to base
the inference on the ensemble behavior instead than on a
single realization. However usually we have just one single
data set. Therefore we propose to use it first to get
information on the statistical distribution over the
ensemble of the $\tilde{\zeta}_j,j=1,\dots,n/2$ which are
the critical quantities  which the parameter estimates are
based on, and then to make use of the data again to get
point and interval estimates of the parameters by a stochastic perturbation method.
Moreover, after the remarks at the end of Section 1, we apply this procedure on different data sets, obtained by dropping some observations at the end of the original data set,  and finally we choose the best result based on a criterium described in the following. For simplicity everywhere - but in Section 1.7 where the proposed algorithm is summarized - we use the same symbol $n$ for the current number of data used.
To describe the procedure is convenient to reformulate
 the original problem  as the one of
estimating the complex measure
$$S(z)=\sum_{j=1}^p c_j\delta(z-\xi_j),\;\;\xi_j\in \mbox{int}(
D), \;\;\xi_j\ne\xi_h \; \forall j\ne h,\;\;c_j\in\C$$
where  $D\subset\C$ is a compact set, from its noisy
moments $$a_k=f(k\Delta)+\epsilon_k,\;k=0,\dots,n-1.$$ It
turns out that \be
s_k=\int_Dz^kS(z)dz=\int\!\!\int_{\!\!\!\!\!\!D}(x+iy)^k
S(x+iy)dx dy, \;\;k=0,1,2,\dots \label{signal} \ee where
\begin{eqnarray}s_k=\sum_{j=1}^p c_j\xi_j^k=f(k\Delta)\label{modale}\end{eqnarray}
hence this problem is equivalent to the original one. We
notice that $S(z)$ is an atomic measure supported on the
(unknown) points $\xi_j,\;j=1,\dots,p$. Estimating a set $\Omega$ such that  $\xi_j\in\Omega,\;j=1,\dots,p,$ is our first goal.

\subsection{The first step}
The idea is  to make use of the relation, discussed in
Section 1, between the numbers $\xi_j,\;j=1,\dots,p$ and
the r.v. $\tilde{\zeta}_j,j=1,\dots,n/2$ which solve the
CEIP for the data
$a_k,\;k=0,\dots,n-1.$ More specifically we want to study
the location in  $\C$ of the $\tilde{\zeta}_j$. As these
are r.v. we are looking for a probability function $h(z)$
defined on the complex plane such that
$$\int_Nh(z)dz=\frac{2}{n}\sum_{k=1}^{n/2}{\mathcal
P}\{\tilde{\zeta}_k\in N\},\;\;N\subset\C.$$ The main
reason to consider the $\tilde{\zeta}_j$ is now apparent:
as $\tilde{\zeta}_j$ are the generalized eigenvalues of the
pencil $U_1(\ua)-zU_0(\ua)$, they are the roots of the
polynomial $Q(z)=\det(U_1(\ua)-zU_0(\ua))$. But then $h(z)$
is the condensed density of these roots  which is given
by (e.g. \cite{j05}):
$$h(z)=\frac{1}{4\pi}\Delta u(z) $$ where $\Delta$ denotes
the Laplacian operator  with respect to $x,y$ if $z=x+iy$
and \begin{eqnarray}
u(z)=\frac{1}{p}E\left\{\log(|Q(z)|^2)\right\}\label{pot}
\end{eqnarray}
is the corresponding logarithmic potential and $E$ is the
expectation operator w.r.to the density of the $a_k$. In
the limit for $\sigma\rightarrow 0$ it can be shown \cite{j08}
that $h(z)$ tends weakly to a measure supported on the
points $\xi_j,\;j=1,\dots,p$. Therefore our first goal is
reached if we are able to compute the expectation in
(\ref{pot}) and to cope with the fact that $h(z)$ conveys
the information on the $\xi_j,\;j=1,\dots,p$ only in the
limit for $\sigma\rightarrow 0$. In \cite{distrf} a closed form
approximation to $h(z)$ based on a single realization
 is provided. The QR decomposition of the random pencil $U_1(\ua)-zU_0(\ua)$ is considered. Then
 $$\log|Q(z)|^2=\sum_{k=1}^{n/2}\log|R_{kk}(z)|^2.$$
 The distribution of $|R_{kk}(z)|^2$ is approximated by a $\Gamma$ density and $u(z)$ is computed analytically.
 Given a realization $\hat{\ua}=\{\hat{a}_k,\;k=0,\dots,n-1\}$ we then get \begin{eqnarray}
\hat{h}(z)\approx\sum_{k=1}^{n/2}\hat{\Delta}
\left(\Psi\left[\frac{1}{2}\left(
\frac{\hat{R}_{kk}^2(z)}{\sigma^2\beta}+1\right)\right]\right)
\label{est} \end{eqnarray}
 where $\hat{\Delta}$ is the discrete
Laplacian evaluated on a square lattice ${\mathcal L}$  of dimension $M$ centered in $(0,0)$ of side greater than one,
$\Psi(x) =\frac{d \log \Gamma(x)}{dx}$ denotes the digamma function,
 $\hat{R}_{kk}^2(z)$ is the diagonal of the $R$ factor in
 the $QR$ factorization of $U_1(\hat{\ua})-zU_0(\hat{\ua})$ and $\beta$
 is an  hyperparameter, to be discussed in the following,
 which control the smoothness of $h(z)$ hence helping in coping with
 the noise. In fact, because of the limit property of
 $h(z)$, if $\sigma$ is small enough there exist disjoint
 sets $N_k,\;k=1,\dots,p$, centered on
 $\xi_k,\;k=1,\dots,p$, such that $\int_Nh(z)dz\approx 1,\;\;N=\bigcup_k
 N_k$. Moreover it was shown in
 \cite{j05,j08} that $h(z)$ can have other noise-related local
 maxima which are located close to the unit circle. However if
 there exist
 signal-related local maxima close to the unit circle they
 can be distinguished from the noise-related ones not only
 by their relative higher magnitude but also by the fact
 that they are surrounded by a set where $h(z)\approx 0$
 (gap of poles of the Pade' approximants \cite{maba2,maba1}).
Increasing $\beta$ will depress the local maxima
  of $h(z)$ and will make larger the sets $N_k$ because
  $h(z)$ is a probability density.
  Eventually some sets $N_k$ will merge together therefore
  determining a loss of resolution but the local
  noise-related maxima will be depressed too and  therefore
  can be easily detected and filtered out by a simple
  thresholding technique which can also make use of
  the "gap of poles" property. Furthermore only a fraction $\tilde{n}=2\tilde{p}<n$ of data are used in this step in order to make an implicit noise filtering. Of course we loose in resolution but this is not relevant in this step.
  Finally we notice that the $QR$ factorization of $U_1(\hat{\ua})-zU_0(\hat{\ua})$ must be computed
for all points of the lattice ${\mathcal L}$. In order to reduce the computational burden,
in \cite{distrf} it was shown that it is enough to compute the $QR$ factorization of the matrix $U$ defined in (\ref{matu}) and then $\forall z$ to upgrade the factorization of $U$ by reducing the Hessemberg matrices $R(E_1-zE_0)$ to triangular form by Givens rotations,
where $E_1$ and $E_0$ are defined in (\ref{e0e1}).

\begin{figure}[h]
\begin{center}
\hspace{1.7cm}\centerline{\fbox{\epsfig{file=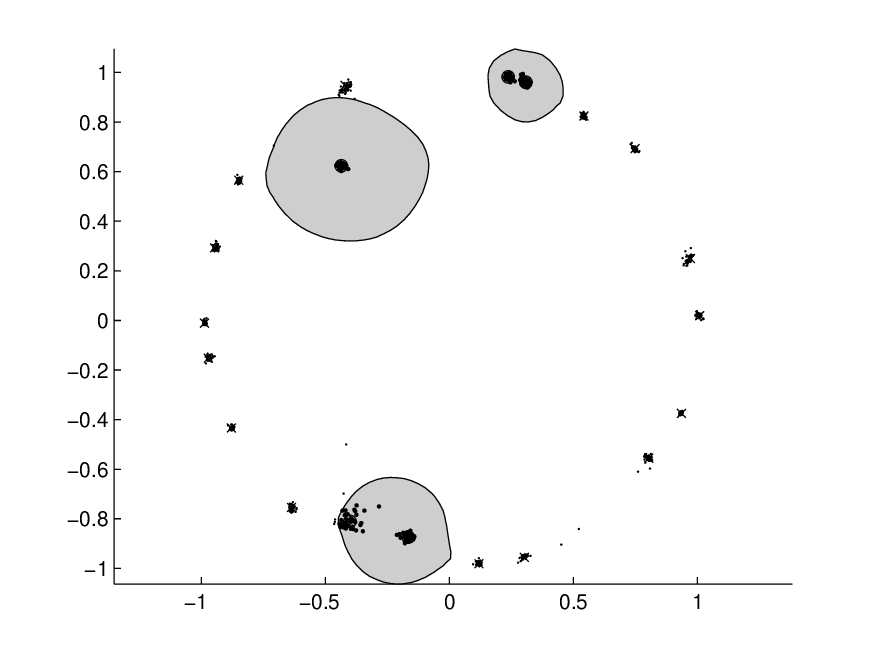,height=4.8cm,width=6cm}}
\hspace{.5cm}\fbox{\epsfig{file=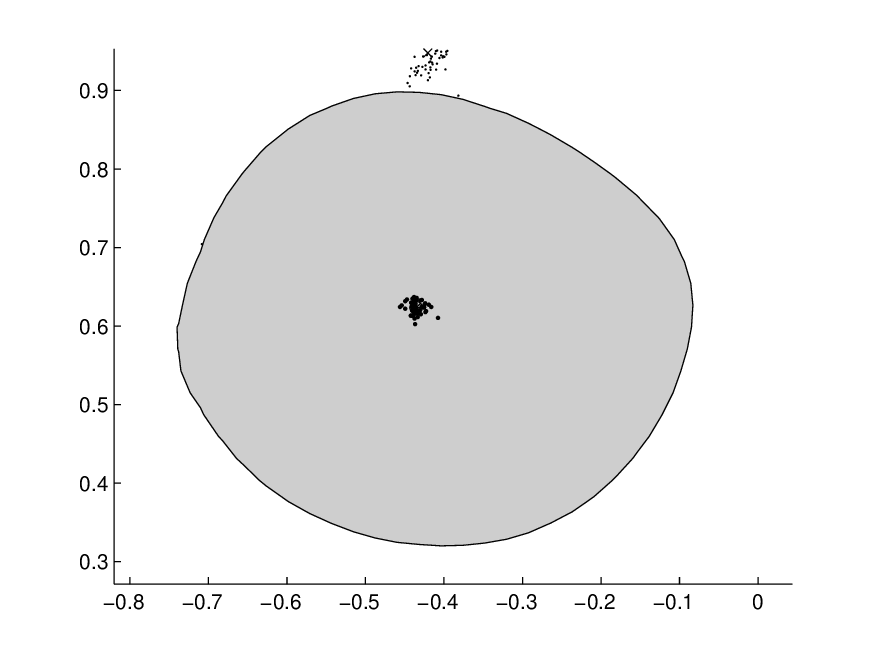,height=4.8cm,width=6cm}}}
\vspace{.2cm}
\hspace{1.7cm}\centerline{\fbox{\epsfig{file=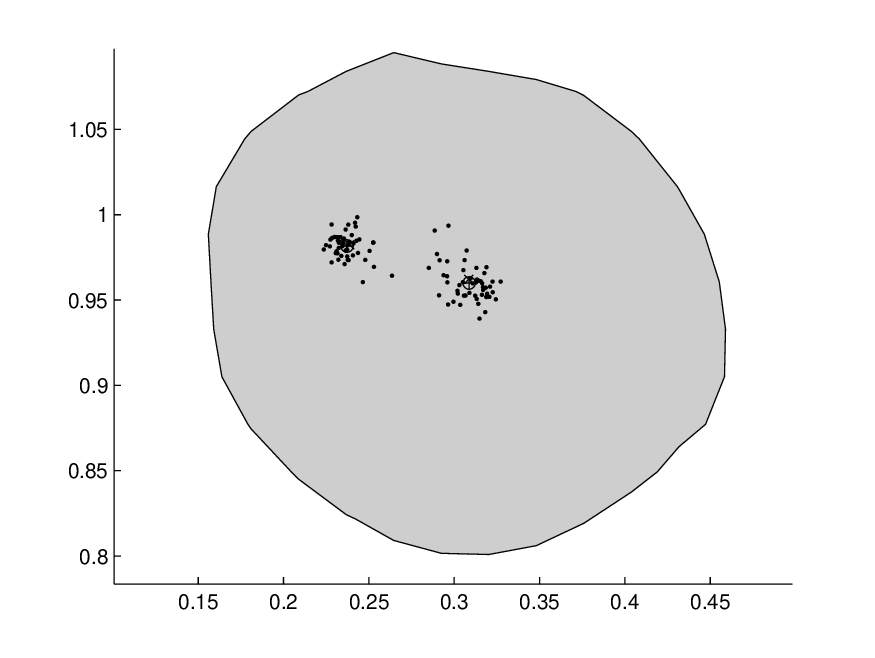,height=4.8cm,width=6cm}}
\hspace{.5cm}\fbox{\epsfig{file=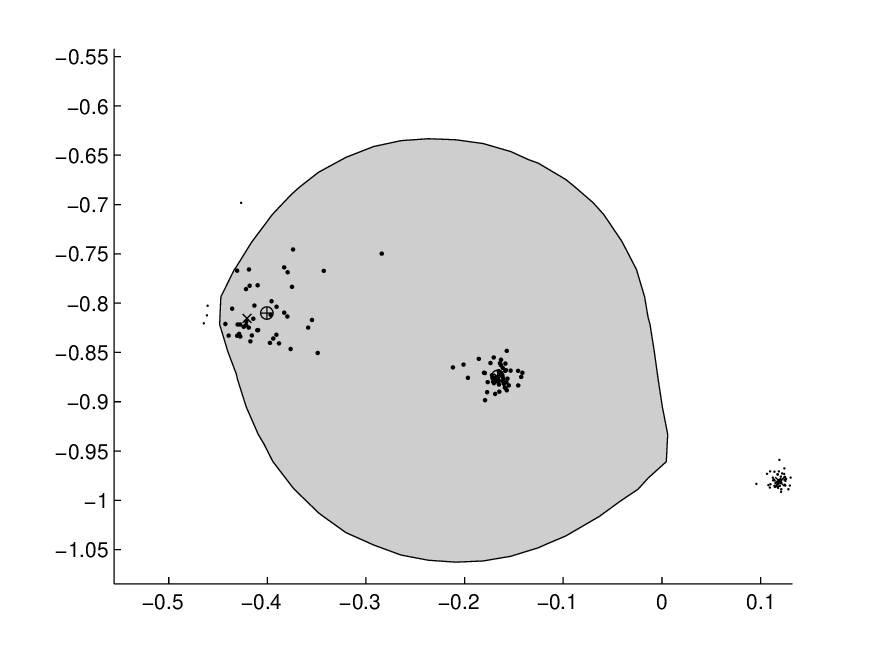,height=4.8cm,width=6cm}}}
\end{center}
\caption{Top left: the sets $N_j,\;j=1,\dots,p_N,\;p_N=3,\; \mbox{ SNR }=0.5$; top right and bottom left and right: zoom of the sets $N_1,N_2,N_3$; the small dots are the generalized eigenvalues corresponding to each pseudosample; the big dots are the generalized eigenvalues falling in $N_1\cup N_2\cup N_3$; the "x" are the initial centroids of the clustering procedure; the "+" are the estimated centroids; the "o" are the centroids of clusters with more than $\alpha\cdot T$ points where $\alpha=0.75$ and $T=300$ is the number of psudosamples. }  \label{lfig6}
\end{figure}

Summing up, in the first step of the procedure the data are
used to identify the sets $N_k,\;k=1,\dots,p_N\le p$ such
that $\xi_j\in N=\bigcup_k
 N_k\;\forall j=1,\dots,p$.
 In fig.\ref{lfig6} top left the results obtained at the end of the first step are shown on a specific example described in Section 3. Three not intersecting sets $N_h$ are computed which contains in their union the true generalized eigenvalues $\xi_k,k=1,\dots,5$.

\subsection{The second step}
  Our second goal is to get
point and interval estimates of the parameters.
  To this purpose a method based on the
 stochastic perturbation idea proposed in \cite{j08} is used.
 Pseudosamples are generated from
 $\{a_k,\;k=0,\dots,n-1\}$ by
 $$a_k^{(r)}=a_k+\nu_k^{(r)},\;\;k=0,\dots,n-1;\;\;\;r=1,\dots,T$$
where $\nu_k^{(r)}$ are i.i.d. zero mean
complex Gaussian variables with variance $\sigma'^2$ independent of
$a_h,\;\forall h$. The CEIP is solved for each of them. GPOF method is used with $n$ data
 and hyperparameters $(l=n/2,\;\tilde{p}).$ The
 generalized eigenvalues are pooled and those not belonging
 to $N$ are discarded.  Then a standard clustering method
 such as e.g. K-means \cite{kmean} is applied to the generalized eigenvalues belonging to $N$ by fixing to $\tilde{p}$ the number of cluster to be estimated and initial centroids given by the solution of the CEIP problem for the $n$ data. The clusters whose cardinality is not close to $T$ are discarded because it was proved in \cite{bama1} that  for each pseudosample it can be expected that in a small neighbor of each $\xi_k,k=1,\dots,p,$ it will fall at least one  estimated generalized eigenvalue. The number of selected clusters
 is an estimate $\hat{p}$ of $p$.
 In fig.\ref{lfig6} top right and bottom left and right
  the big dots indicates the generalized eigenvalues  which belong to  $N$ on a specific case and small dots indicates the generalized eigenvalues  which do not belong to $N$. We notice the presence of several spurious clusters of generalized eigenvalues which justify the importance of the first step of the procedure.
 The
 estimates $\hat{\xi}_k$ of $\xi_k$ are then computed by averaging the generalized eigenvalues belonging to the $k-$th
cluster. The estimates $\hat{c}_k$ of $c_k$ are then
computed by solving the standard  least squares problem
$$\hat{\uc}=\argmin_{\uga}\|V(\hat{\uxi})\uga-\ua\|^2.$$
We notice  that interval estimates of $\uxi$ and $\uc$ can also be obtained from the clustering results.

\subsection{Estimation of $\beta$}

The first step of the procedure depends critically on the
choice of $\beta$. A value of $\beta$ too small will give
rise to many modes of $h(z)$  which are likely to be
spurious  but not easily detectable as noise-related ones.
A value of $\beta$ too large will give rise to a small
number of modes, possibly much less than $p$. The
clustering method can then become critical. The idea for
getting a good value for $\beta$ is based on a comparison
of formula (\ref{est}) with another approximation of $h(z)$
  given in \cite{j08} by:
$$\tilde{h}(z)=\frac{1}{2\pi n}\Delta\sum_{\mu_j(z)> 0
}\log(\mu_j(z))$$ where $\mu_j(z)$ are the eigenvalues of
 \be (U_1(\us)-zU_0(\us))\overline{(U_1(\us)-zU_0(\us))}+
\frac{n\sigma^2}{2}A(z,\overline{z})\label{effetton}\ee
where $A(z,\overline{z})\in \C^{n/2\times n/2}$ is a
tridiagonal hermitian matrix with $1+|z|^2$ on the leading
diagonal and $-\overline{z}$ and $-z$ on the diagonals
respectively below and above the leading one. As the components of the vector
$\us$ given in (\ref{signal}) are unknown, this
formula cannot be used to estimate $h(z)$. However we notice that
$\tilde{h}(z;\sigma))=\frac{1}{2\pi n}\Delta\log\det(UU^H+\frac{n}{2}\sigma^2 A)$ where $U=U_1(\us)-zU_0(\us)$. Let  $U=QR$ be the $QR$ decomposition of $U$ where the diagonal of $R$ can be assumed to be real. As $U=U^T$ we also have $U=R^TQ^T$ and therefore
$UU^H=R^TQ^T\overline{Q}\overline{R}=R^T\overline{R}$ because $Q$ is unitary. But then
\begin{equation}\tilde{h}(z;\sigma))=\frac{1}{2\pi n}\Delta\log\det(R^HR+\frac{n}{2}\sigma^2 \overline{A}).\label{appr1}\end{equation}
We notice that formula (\ref{est}) is an approximation of
(see \cite[eq.6]{distrf})

\begin{eqnarray}\hat{h}(z;\sigma,\beta) & = & \frac{1}{2\pi n}\Delta\sum_{k=1}^{n/2}
\left(\Psi\left[\frac{1}{2}\left(
\frac{E[\hat{R}_{kk}^2]}{\sigma^2\beta}+1\right)\right]\right) \nonumber \\& \approx &
\frac{1}{2\pi n}\Delta\sum_{k=1}^{n/2}
\left(\log\left[\frac{1}{2}\left(
\frac{E[\hat{R}_{kk}^2]}{\sigma^2\beta}+1\right)\right]\right) \nonumber \\
&=&\frac{1}{2\pi n}\Delta\sum_{k=1}^{n/2}
\log\left[E[\hat{R}_{kk}^2]+\sigma^2\beta\right] \label{appr2}\end{eqnarray}
where $\hat{R}_{kk}^2(z)$ is the diagonal of the $R$ factor in
 the $QR$ factorization of $U_1(\ua)-zU_0(\ua).$
Therefore we can compare formula (\ref{appr2})  with formula (\ref{appr1}).
Let us assume that $E[\hat{R}_{kk}^2]\approx R_{kk}^2$ and consider the case when $z=0$.
Formula (\ref{appr1}) and formula (\ref{appr2})  become respectively
$$\tilde{h}(z;\sigma))\approx\frac{1}{2\pi n}\Delta\log\det(R^HR+\frac{n}{2}\sigma^2 I)$$
and
$$ \hat{h}(z;\sigma,\beta)=\frac{1}{2\pi n}\Delta\sum_{k=1}^{n/2}
\log\left[R_{kk}^2+\sigma^2\beta\right].$$
As $\log[\det(R^H R)]=\sum_k\log( R_{kk}^2)$, $\tilde{h}(z;\sigma))$ and $\hat{h}(z;\sigma,\beta)$  are close when $\beta=\frac{n}{2}$ and $\sigma\rightarrow 0$ or when $R$ is a diagonal matrix.
This suggests to use $\frac{n}{2}$ as the initial guess for $\beta$ and then to increase it by a little amount to get a smoother estimate of $h(z)$ useful for estimating the set $\Omega$ in the first step. In the following the value $\beta=\beta_a\tilde{p}$ is used where $\beta_a\ge 1$ is an amplification factor .

\subsection{Filtering the QR decomposition}
It turns out that the first step of the procedure depends critically on the QR factorization of the matrix $U_1(\ua)-zU_0(\ua)$ or, as proved in \cite{distrf}, on that of the matrix $U$ defined in (\ref{matu}). It is therefore necessary to filter out the noise from the $R$ factor of $U$. This is a very delicate task which can be however successfully  accomplished by taking into account the special structure of the data as follows. We notice that the real and imaginary parts of the signal $f(t) = \sum_{j=1}^pc_j\xi_j^t$ decay to zero exponentially.
However when Gaussian noise is present the tail of the data fill a rectangular region centered on the $t-$axis of width  $\approx 2\sqrt{2}\sigma$. A classic way to reduce the contribution of the noise consists therefore in applying an exponential filter to force the tail of the data to go to zero as in the noiseless case. In section 2.4 we discussed the Cadzow iteration to filter out the noise in $U$ without destroying its Hankel structure. However, in order to further improve the  estimate of $R$,  we suggest to apply a filter also after the factorization process.

\begin{figure}
\begin{center}
\hspace{1.7cm}{\fbox{\psfig{file=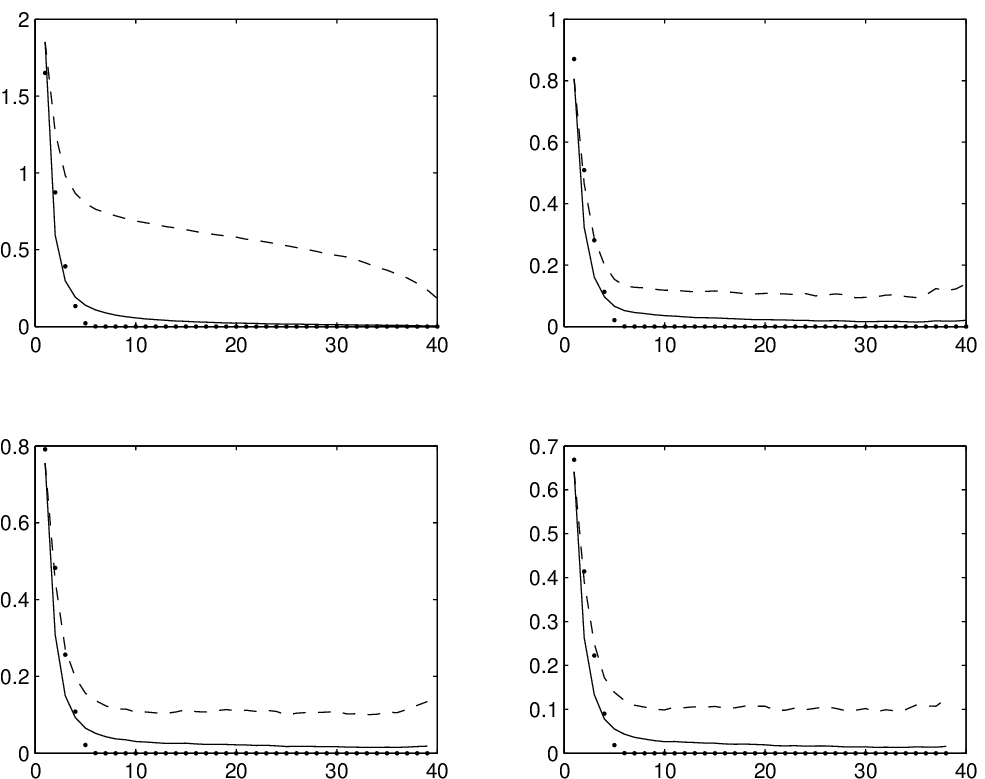,height=16cm,width=16cm}}}
\end{center}
\caption{  Top right: the main diagonal of the matrix $R$ in the noiseless case (dotted), in the noisy case (dashed) and the filtered one (solid) are represented when SNR$=1,\;\;p=5$. Top left: the same for the absolute value of the first diagonal. Bottom left: the same for the absolute value of the second diagonal. Bottom right: the same for the  absolute value of the third diagonal.}\label{lfiltro}
\end{figure}

To this aim
we notice first that, if $U=QR,\;Q^HQ=I,\;\;R\mbox{  upper trapezoidal}$, the main diagonal of $R$ can be chosen to be non-negative and monotonic decreasing. In the noiseless case the last $n-p$ rows of $R$ must be zero, as $\mbox{rank}(U)=p$. It can be shown experimentally that the same behavior characterizes also the absolute value of the secondary diagonals $\left\{|R_{h,h+l}|,\;h=1,\dots,p-l\right\},\;\;l=0,\dots,p-1$. Moreover this behavior  is preserved also in the noisy case but with an asymptotic value greater than zero. In fig.\ref{lfiltro} the results of a simulation showing these facts are reported. A set of  complex exponential signals were generated with random  frequencies $\omega_j$ and phases $\theta_j$ with uniform distribution in $[-\pi,\pi)$, random decays $\rho_j$ with uniform distribution in $(0,1]$ and complex standard Gaussian random amplitudes normalized in order to make their absolute values to sum to one. The matrix $U$ was then formed and the QR decomposition was computed. The absolute values of the diagonals of $R$ were then averaged and the results for the main diagonal and the first three secondary diagonals was plotted. The same is done by adding complex Gaussian white noise to the complex exponential signals.

The comparison of the results in the noiseless and noisy cases for several SNRs and orders $\tilde{p}$, suggests that we can filter out the noise in the diagonals of $R$ by $$\tilde{R}_{h,h+l}=\frac{R_{h,h+l}}{h^{\gamma_l}},\;h=1,\dots,\tilde{p}-l ,\;\;\gamma_l>0,\;\;l=0,\dots,\tilde{p}-1.$$
In fig(\ref{lfiltro}) the filtered diagonals were plotted too where $\gamma$ was estimated by solving the problems $$\hat{\gamma}_l=\mbox{argmin}_\gamma \sum_{h=1}^{\tilde{p}-l}|\tilde{R}_{h,h+l}-R_{h,h+l}|,\;\;l=0,\dots,\tilde{p}-1.$$ It can be noticed a good agreement between the noiseless and filtered data, therefore the functional form of the filter seems to be adequate to do the job. In the following we  choose only one hyperparameter $\gamma$ and filter the diagonals of $R$ according to the rule
\begin{eqnarray}\tilde{R}_{h,h+l}=
 \frac{R_{h,h+l}}{h^\gamma}
 \label{filtro}.
\end{eqnarray}

\subsection{The criterium for choosing the data set}

Up to now we have considered the number of data $n$ as fixed. From the remarks at the end of Section 1, we know that $n$  is a critical parameter. Therefore we want to choose it in an optimal way. Let us assume that the given number of data $n_{orig}$ is such that  $f(n_{orig}\Delta;p,P)\approx 0.$ In Section 1 we have conjectured that dropping some data at the end of the original data set could decrease the bias of the estimator of the parameters $\xi_j,\;j=1,\dots,p$. By hypothesis we know that the noise affecting the data is i.i.d., therefore the residuals corresponding to the true parameters $P$ will form a stationary sequence i.e. their autocorrelation function will be a Kronecker $\delta$ sequence. If we consider the residuals corresponding to the estimated parameters we can expect that some signal component is still present in the residuals and therefore the autocorrelation function will be different from zero for some lag greater than zero. We can then use the following statistics to quantify the goodness of the estimation as a function of the number $n\le n_{orig}$ of used data:
$$C(n)=\frac{2}{n R(0)^2}\sum_{k=1}^{n/2}|R(k)|^2$$ where $$R(k)=\sum_{h=0}^{n-k-1} (\hat{\epsilon}_{h+k} -\hat{\mu}) (\overline{\hat{\epsilon}_{h}-\hat{\mu}}),\;\;\hat{\epsilon}_h=a_h-f(h\Delta;\hat{p},\hat{P}),\;\;
\hat{\mu}=\frac{1}{n}\sum_{h=0}^{n-1}\hat{\epsilon}_h .$$
The optimal $n$ will be
$$n_{ott}=\mbox{argmin}_n C(n).$$
We notice that when $\hat{P}=P$ then $C(n)=0 \;\;\forall n$ and the dependence on $n$ of $C(n)$ is only through $\hat{P}$ because of the division by $n$ in the definition.

\subsection{The algorithm}
Summing up, a sketch of the proposed algorithm is the following:
\begin{itemize}
\item fix a square lattice ${\mathcal L}$  of dimension $M$ centered in $(0,0)$ of side $L>1$
\item fix an initial even number $n_0<n_{orig}$ of data such that $n_0/2\gg p$ and an estimate $\hat{\sigma}$ of $\sigma$
\item for  $n=n_0+k\Delta_n,\;k=0,\dots,K$ and $\Delta_n\in\N^+$, even,  such that $K=\left\lfloor\frac{n_{orig}-n_0}{\Delta_n}\right\rfloor$
\begin{itemize}
\item[-] choose the number $\tilde{p}$ of signal-related components as a fixed percentage of the current number of data $n$
\item[-] compute $U$ based on the first $n$ data and filter it by Cadzow algorithm using $l=n/2$ and $\tilde{p}$, producing $n$ filtered data
\item[-] compute $U=QR$ based on the first $2 \tilde{p}$ filtered data and filter the diagonals of $R$ by formula (\ref{filtro})
\item[-] compute the Hessemberg matrices $R(E_1-zE_0),\;\forall z\in{\mathcal L}$  and reduce them to triangular form by Givens rotations
\item[-] compute $\hat{h}(z;\beta),\;\beta=\beta_a\tilde{p},$ by formula (\ref{est}) where $\hat{R}_{kk}(z)$ are the diagonal elements of the triangular matrices computed in the previous step
\item[-] compute the  sets $N_k,\;k=1,\dots,p_N$  such that
\begin{itemize}
  \item[$\circ$] $\hat{h}(z;\beta)$ is unimodal for $z\in N_k$
  \item[$\circ$]  $ \bigcap_{k=1}^{p_N}N_k=\emptyset$
 \end{itemize}
 by selecting the local maxima of $\hat{h}(z;\beta)$ above a given threshold $\tau>0$, and then
 by identifying the neighbor $N_k$ of the k-th local maxima $\hat{\xi}_k$ such that $\hat{h}(z;\beta)$ is monotonic decreasing along the four coordinate directions on the lattice ${\mathcal L}$ starting from $\hat{\xi}_k$
\item[-] generate $T$ pseudosamples based on the filtered $n$ data
\item[-] solve the CEIP for each pseudosample by GPOF method with hyperparameters $l=n/2,\tilde{p}$ and pool the $\xi^{(r)}_h$
\item[-] cluster the $\xi^{(r)}_h\in \bigcup N_k$ and discard the others. The   k-means method is used to find $\tilde{p}$  clusters;  the clusters with less than $\lfloor\alpha T\rfloor,\;\;\alpha\in(0.5,1]$ elements are discarded
\item[-] $p_{ott}(n)=$ number of selected clusters
\item[-] $\hat{\xi}_k(n)\;=$ average of the $ \xi^{(r)}_h($ in  cluster $k$-th, $k=1,\dots,p_{ott}(n)$
\item[-] $\hat{c}_k(n)\;=$ average of the $ c^{(r)}_h$ in  cluster $k$-th, $k=1,\dots,p_{ott}(n)$
\item[-]  compute $C(n)$ and memorize $\hat{\xi}_k(n),\hat{c}_k(n),\;k=1,\dots,p_{ott}(n)$
\end{itemize}
\item compute $n_{ott}=\mbox{argmin}_n C(n).$ The optimal parameter estimates are
$$\hat{\xi}_k(n_{ott}),\;\;\hat{c}_k(n_{ott}),\;\;k=1,\dots,p_{ott}(n_{ott})$$

\end{itemize}
The required hyperparameters are: the lattice dimension $M$, the side of the lattice $L$,  the step $\Delta_n$ for choosing the current number of data, the number $\nu$ of iterations of the Cadzow algorithm, the amplification factor $\beta_a$ for the smoothing parameter $\beta=\beta_a\tilde{p}$, the filter parameter $\gamma$ for the diagonals of $R$,  the threshold $\tau$ for selecting the local maxima of the condensed density, the ratio between $\tilde{p}$ and the maximum number $n/2$ of estimable components, the number of pseudosamples $T$, the ratio between standard deviation of pseudosamples and noise standard deviation $\frac{\sigma'}{\sigma}$, the acceptation threshold for clusters $\alpha$.

Also the noise variance $\sigma^2$ and the initial number of data $n_0$ could be considered as  hyperparameters. However there turns out that they are the only hyperparameters which are data dependent.
If enough data are measured in order that the signal is decayed below the noise threshold  then an estimate of $\sigma$ can be obtained from the tail of the data.
For choosing a good value of $n_0$
the following considerations can help. As the criterium $C(n)$ for choosing the optimal data set measures the stationarity of the residuals, if in the true signal there are components $c_j\xi_j^t$ much smaller than others (e.g. with respect to the $L_2$ norm $\int_0^\infty|c_j\xi_j^t|^2dt$)  there can happen that $C(n_1)<C(n_2)$ and $n_1<n_2$ but the small components are missed if $n_1$ data are used. In these cases it is not convenient to choose a small value of $n_0$. On the contrary if $n_0$ is too large, less degree of freedom are left to the procedure for choosing the best data set and therefore a poor estimation could result but  when the components are all close to pure sinusoids. In fact in this case the quality of the estimation improves by increasing $n$ as shown in Section 1.3 and fig.\ref{lfig0}.
It seems therefore reasonable to leave some flexibility to the proposed black box method by letting the user to choose $\sigma$ and $n_0$.

\section{Simulation results}

 In order to test the advantages of the proposed method w.r.to the standard ones, four experiments were performed corresponding to the four values of the noise s.d.
$\sigma=2\sqrt{2},\sqrt{2},\frac{\sqrt{2}}{3},\frac{\sqrt{2}}{10}.$ In each experiment
$N=300$ independent
  realizations of the r.v. $a_k^{(h)},k=1,\dots,n_{orig}=120,\;\;h=1,\dots,N$ were
  generated from the complex exponentials model with $p=5$ components given by
$$\underline{\xi}=\left[e^{-0.3-i 2\pi
0.35}, e^{-0.1-i 2\pi  0.3},e^{-0.05-i 2\pi
0.28},e^{-0.0001+i 2\pi 0.2},e^{-0.0001+i 2\pi  0.21}\right]$$ $$ \underline{c}=\left[ 20,6,3,1,1\right
]$$ by adding complex Gaussian noise with s.d. $\sigma$.
  We notice that  the frequencies of the $4^{rd}$
and $5^{th}$ components are closer than the Nyquist frequency
if $n<1/(0.21-0.20)=100$.  By defining $SNR_i=\sqrt{2}\frac{|c_i|}{\sigma}$ we label
 the four considered cases by $SNR=\min_i SNR_i=[0.5,1,3,10]$. The choice $N=300$ makes simulation results almost independent of the initialization of the pseudorandom numbers generator.
\noindent For each experiment and for each $h=1,\dots,N$ the method GPOF \cite{hs0} was applied. After some trials the best results were obtained by using the hyperparameters
 $l=m/2,\;\;m=n_{orig}/2,\dots,n_{orig}$ and $\tilde{p}=1,\dots,l/2$. For each estimate $\uzet(m,\tilde{p})$ of the generalized eigenvalues, the corresponding  estimates $\gamma(m,\tilde{p})$ of the weights was obtained by solving a linear least squares problem.
 For each fixed dataset $a_k^{(h)},k=1,\dots,m$ the optimal model order $p_o(m)$ was chosen by minimizing the BIC criterium \cite{aka} as a function of $\tilde{p}$. BIC was used because it provides the best results in this framework among AIC, $\mbox{AIC}_c$ and SIC (see e.g.\cite{burnand} for the definition of these criteria). The optimal dataset $a_k^{(h)},k=1,\dots,m_{ott}$ was determined by minimizing the residual stationarity criterium considered as a function of  $m$ and $p_o(m)$, i.e. $m_{ott}=\min_m C(m,p_o(m))$ and therefore the optimal model order was given by  $p_{ott}=p_o(m_{ott})$.
 The corresponding optimal parameters
$\hat{\xi}_j$ and $\hat{c}_j$  were  computed. $|\hat{c}_j|$ were then sorted in descending order and  $\hat{\xi}_j$ were sorted accordingly. $\hat{c}_j$ and $\hat{\xi}_j$ were then used to estimate the signal by $$\hat{s}_k=\sum_{j=1}^{p_{ott}}\hat{c}_j\hat{\xi}_j^k.$$
If $p_{ott}\ge p$, the relative error was computed by
$$E(\sigma,h)=\frac{\sum_{j=1}^p|c_j-\hat{c}_j|^2}{\sum_{j=1}^p |c_j|^2}+
\frac{\sum_{j=1}^p|\xi_j-\hat{\xi}_j|^2}{\sum_{j=1}^p|\xi_j|^2},\;\;h=1,\dots,N.$$ Otherwise
$E(\sigma,h)$ was set to the conventional value $-1$.
  The average relative MSEs
$$MSE(\sigma)=\frac{1}{N_\sigma}\sum_{h=1}^{N_\sigma} E(\sigma,h)$$
where $N_\sigma$ is the cardinality of the set $\{h|E(\sigma,h)\ge 0\}$,
are reported in the first row of Table \ref{tb1}. The values of relative MSEs greater than one indicates that even if a sufficient number of components has been identified, at least some of them are poorly estimated. In the second row the cardinalities $N_\sigma$ are reported.
\begin{table}[tbh]
\begin{scriptsize}
\begin{center}
\begin{tabular}{|c|c|c|c|c|}
\hline\hline
&$SNR=0.5$&$SNR=1$&SNR=3&SNR=10\\
\hline
$MSE(standard)$&1.302&0.865&0.310&0.095\\
\hline
$N_\sigma$&173&123&220&300\\
\hline\hline
$MSE(proposed)$&0.860&0.635&0.283&0.136\\
\hline
$N_\sigma$&100&201&285&300\\
\hline
\end{tabular}
\end{center}
\caption{Standard method: relative MSEs (first row) averaged over $N_\sigma$ (second row) replications. Proposed method: relative MSEs (third row) averaged over $N_\sigma$ (fourth row) replications. } \label{tb1}
\end{scriptsize}
\end{table}

\noindent The new method was then applied to the same data. The algorithm illustrated in section 2.7 was applied with  $n_0=30 $, $\sigma=2\sqrt{2},\sqrt{2},\frac{\sqrt{2}}{3},\frac{\sqrt{2}}{10}$ and $h=1,\dots,N.$  The numerical values of the hyperparameters used in all the experiments reported in this section are given in Table \ref{tb0}. They were obtained once and for all by trials and errors on one data set with SNR$=1$ stopping the search when better results  than those provided by the standard method were obtained. The search was not pursued further because we want to study the average behavior of the proposed method and its robustness with respect to  the hyperparameters. It is therefore possible that a fine tuning of the hyperparmeters can improve the results on specific instances.

\begin{table}[tbh]
\begin{center}
\begin{scriptsize}
\begin{tabular}{|c|c|c|c|c|c|c|c|c|c|c|}
\hline\hline
$ M$ & $L$ & $\Delta_n$ & $\nu$ & $\beta_a$ & $\gamma$ & $\tau$ & $\frac{2\tilde{p}}{n}$ & $T$ & $\frac{\sigma'}{\sigma}$ & $\alpha$ \\
\hline\hline
$80$  & $1.1$ &$10$ & $10$ & $1.2$ & $0.4$ & $0.002$ & $0.3$ & $30$ & $0.15$ & $0.75$ \\
\hline \hline
\end{tabular}
\caption{Hyperparameters. }
\label{tb0}
\end{scriptsize}
\end{center}
\end{table}

\begin{figure}[h]
\begin{center}
\hspace{1.7cm}\centerline{\fbox{\epsfig{file=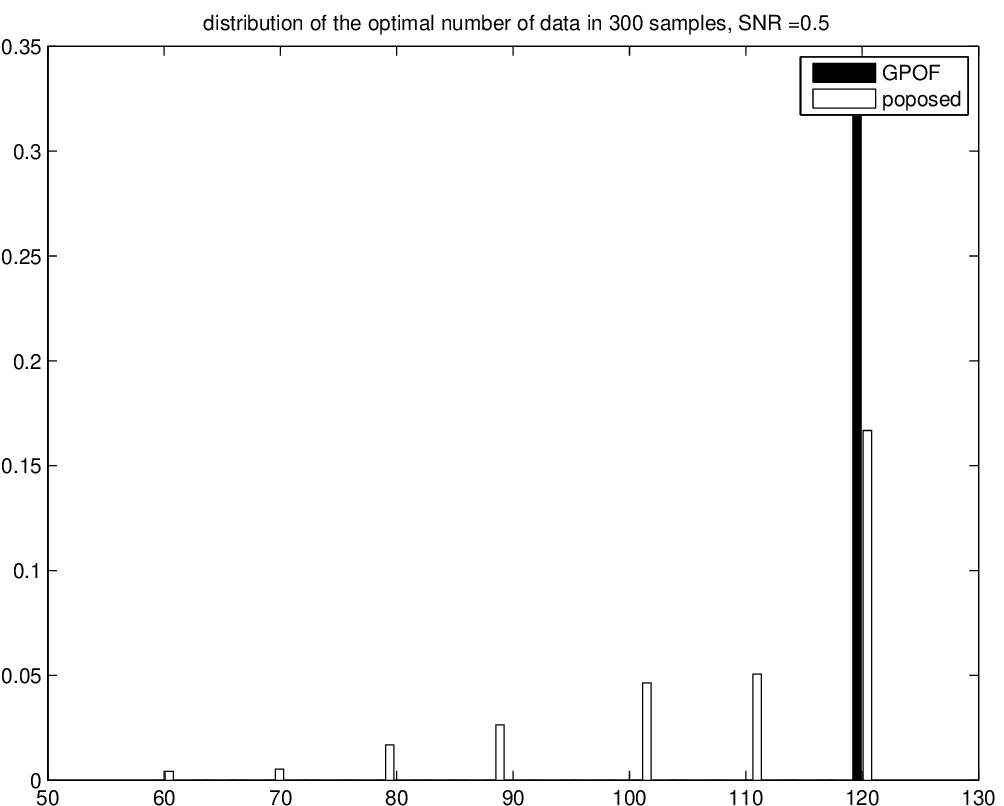,height=4.8cm,width=6cm}}
\hspace{.5cm}\fbox{\epsfig{file=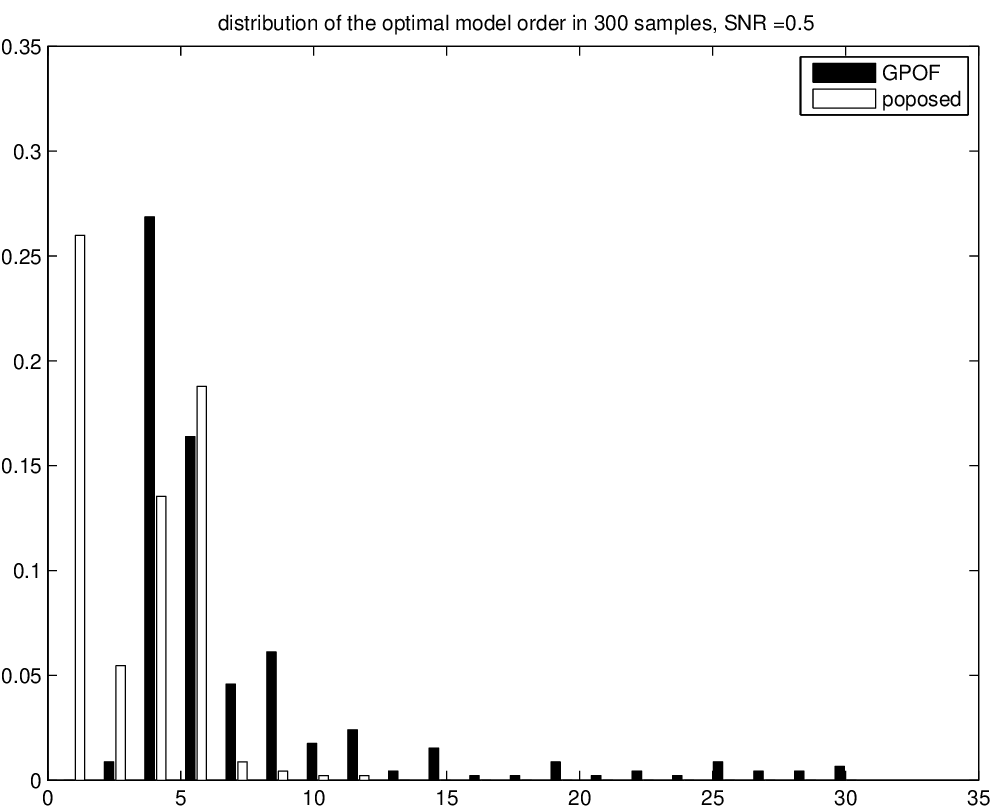,height=4.8cm,width=6cm}}}
\vspace{.2cm}
\hspace{1.8cm}\centerline{\fbox{\epsfig{file=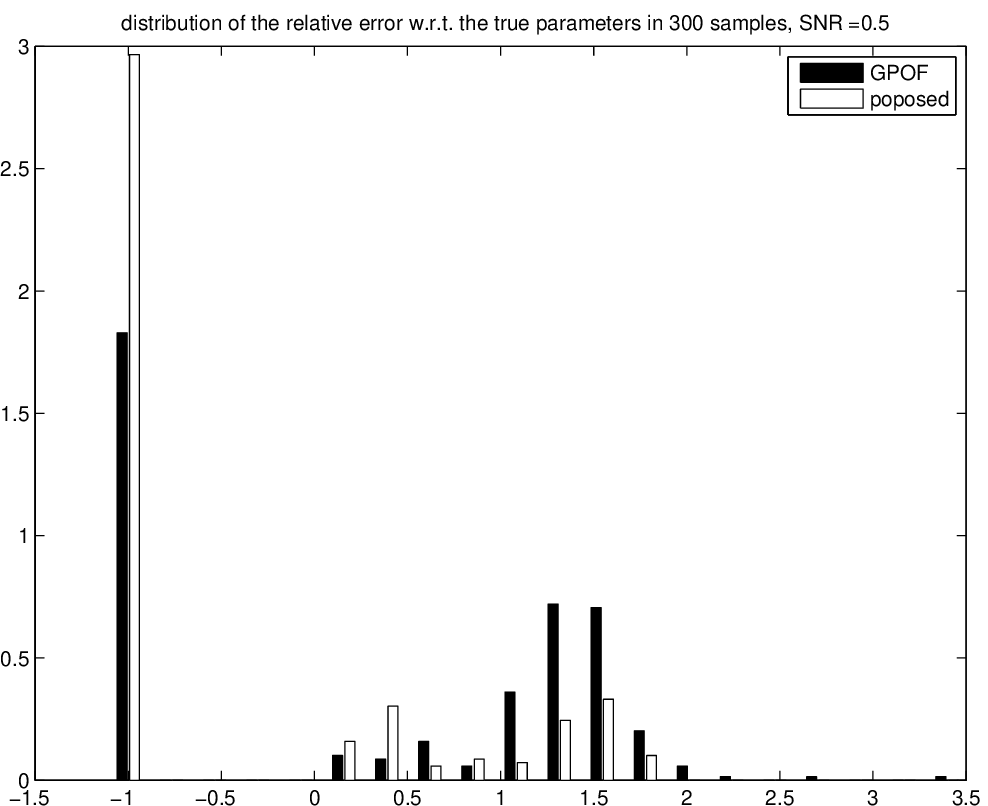,height=10cm,width=12.8cm}}
}
\end{center}
\caption{The empirical distributions over the $300$ replications of $n_{ott}(\sigma,\cdot)$ (top left), $p_{ott}(\sigma,\cdot)$ (top right) and $E(\sigma,\cdot)$ (bottom) for $\sigma=2\sqrt{2}$. The class $-1$ represents the samples where the true number of components was underestimated.  }  \label{lfig71}
\end{figure}

\begin{figure}[h]
\begin{center}
\hspace{1.7cm}\centerline{\fbox{\epsfig{file=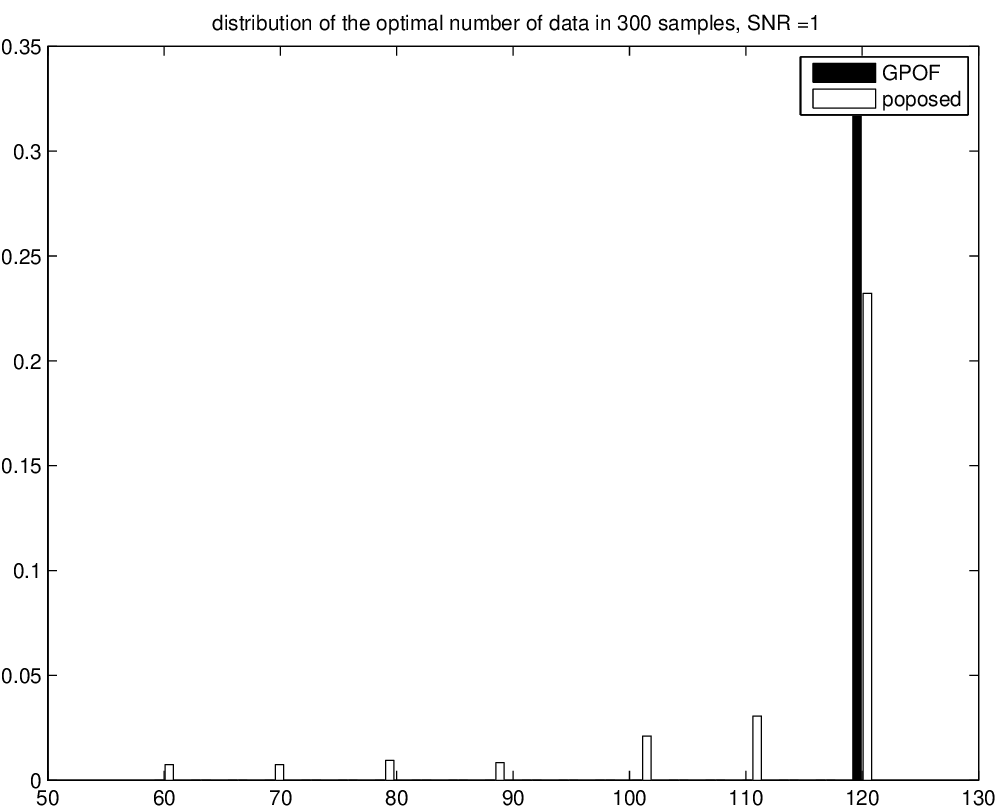,height=4.8cm,width=6cm}}
\hspace{.5cm}\fbox{\epsfig{file=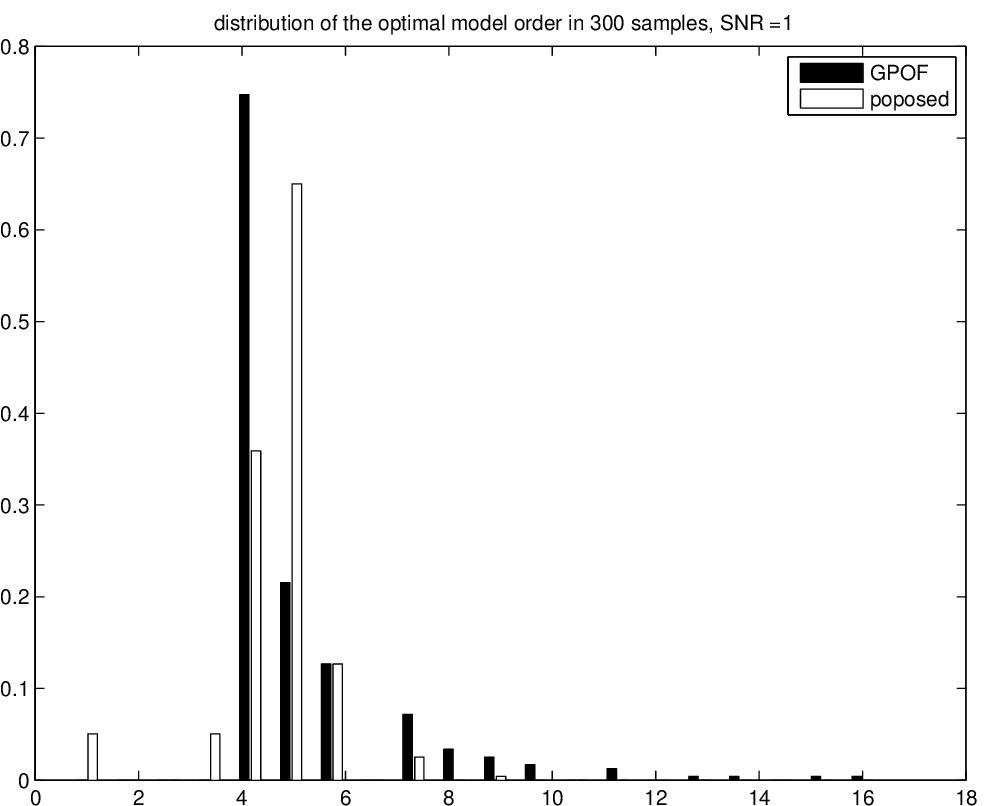,height=4.8cm,width=6cm}}}
\vspace{.2cm}
\hspace{1.8cm}\centerline{\fbox{\epsfig{file=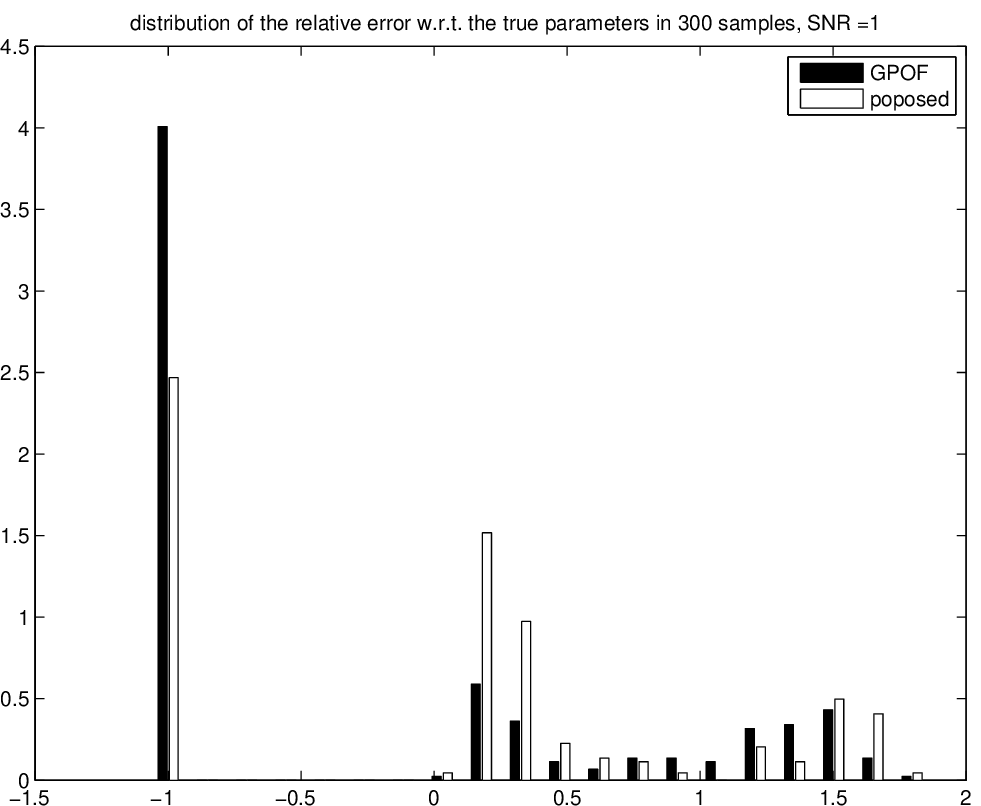,height=10cm,width=12.8cm}}
}
\end{center}
\caption{The empirical distributions over the $300$ replications of $n_{ott}(\sigma,\cdot)$ (top left), $p_{ott}(\sigma,\cdot)$ (top right) and $E(\sigma,\cdot)$ (bottom) for $\sigma=\sqrt{2}$. The class $-1$ represents the samples where the true number of components was underestimated.  }  \label{lfig72}
\end{figure}

\begin{figure}[h]
\begin{center}
\hspace{1.7cm}\centerline{\fbox{\epsfig{file=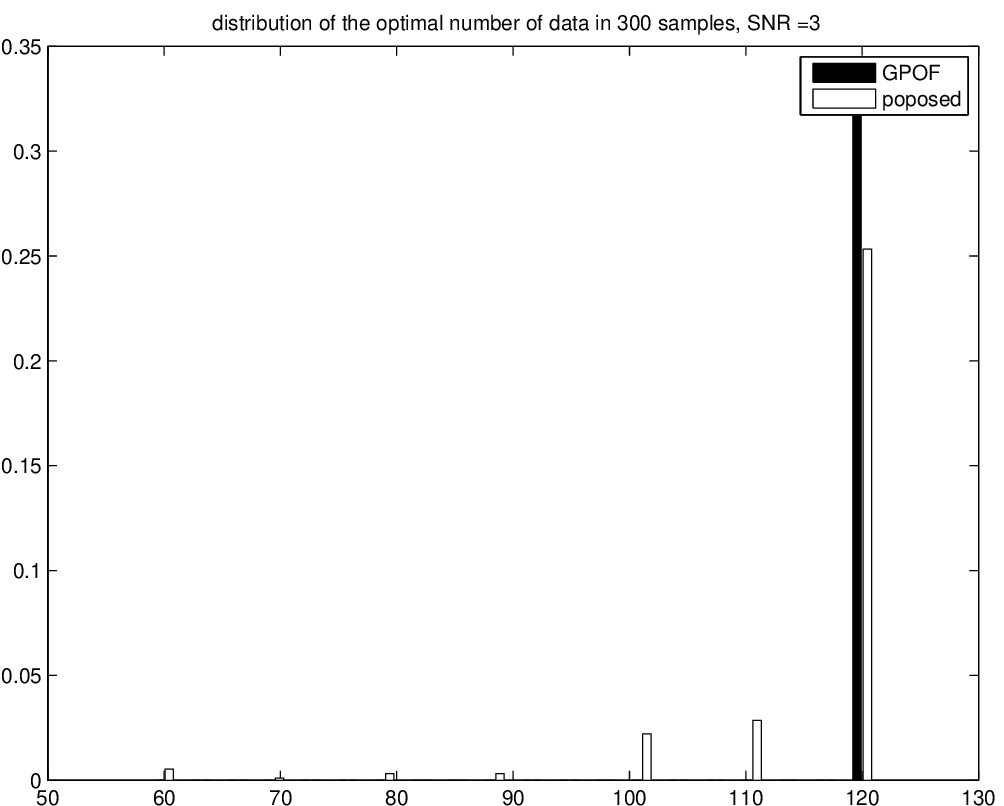,height=4.8cm,width=6cm}}
\hspace{.5cm}\fbox{\epsfig{file=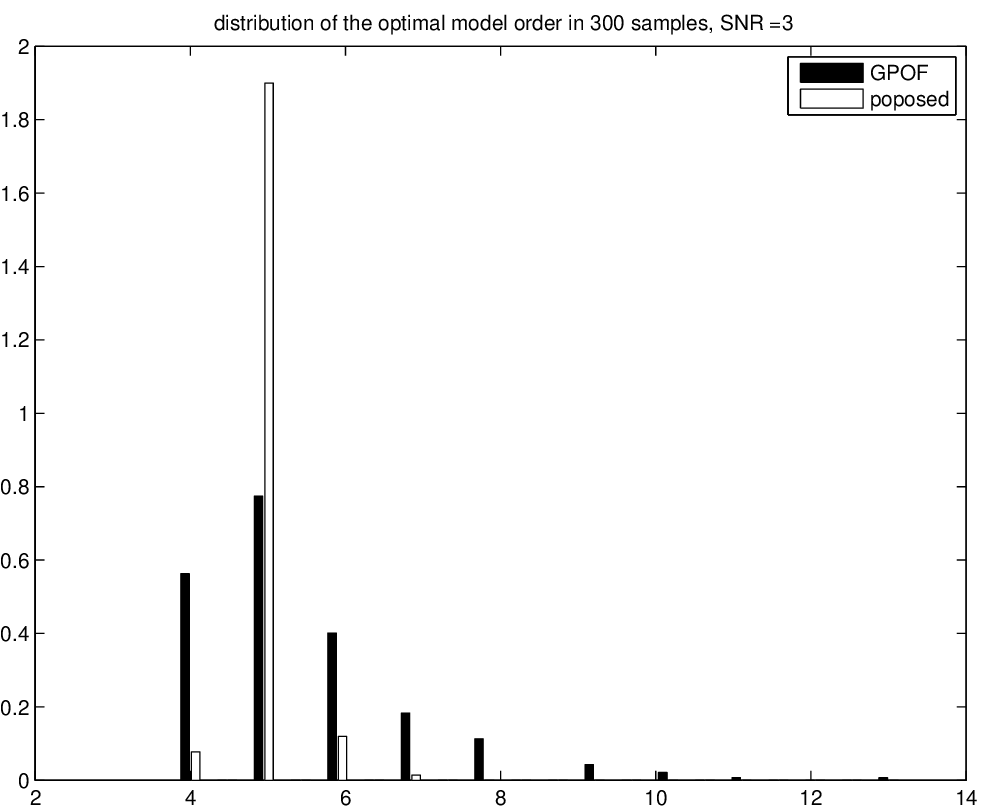,height=4.8cm,width=6cm}}}
\vspace{.2cm}
\hspace{1.8cm}\centerline{\fbox{\epsfig{file=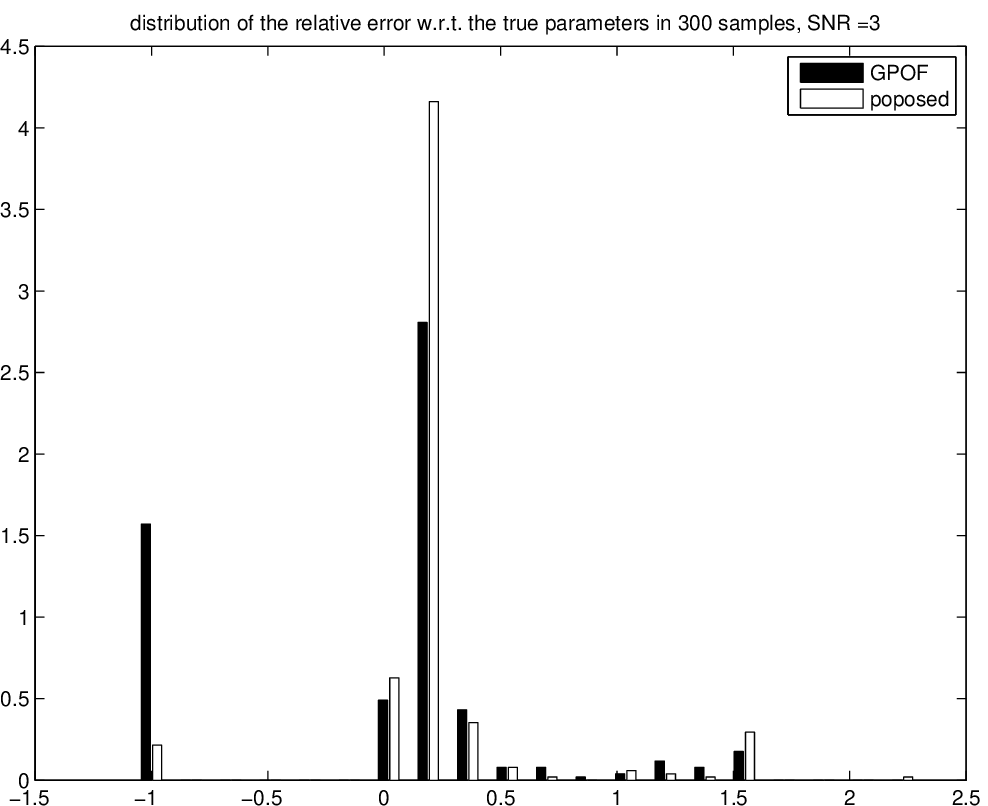,height=10cm,width=12.8cm}}
}
\end{center}
\caption{The empirical distributions over the $300$ replications of $n_{ott}(\sigma,\cdot)$ (top left), $p_{ott}(\sigma,\cdot)$ (top right) and $E(\sigma,\cdot)$ (bottom) for $\sigma=\sqrt{2}/3$. The class $-1$ represents the samples where the true number of components was underestimated.  }  \label{lfig73}
\end{figure}

\begin{figure}[h]
\begin{center}
\hspace{1.7cm}\centerline{\fbox{\epsfig{file=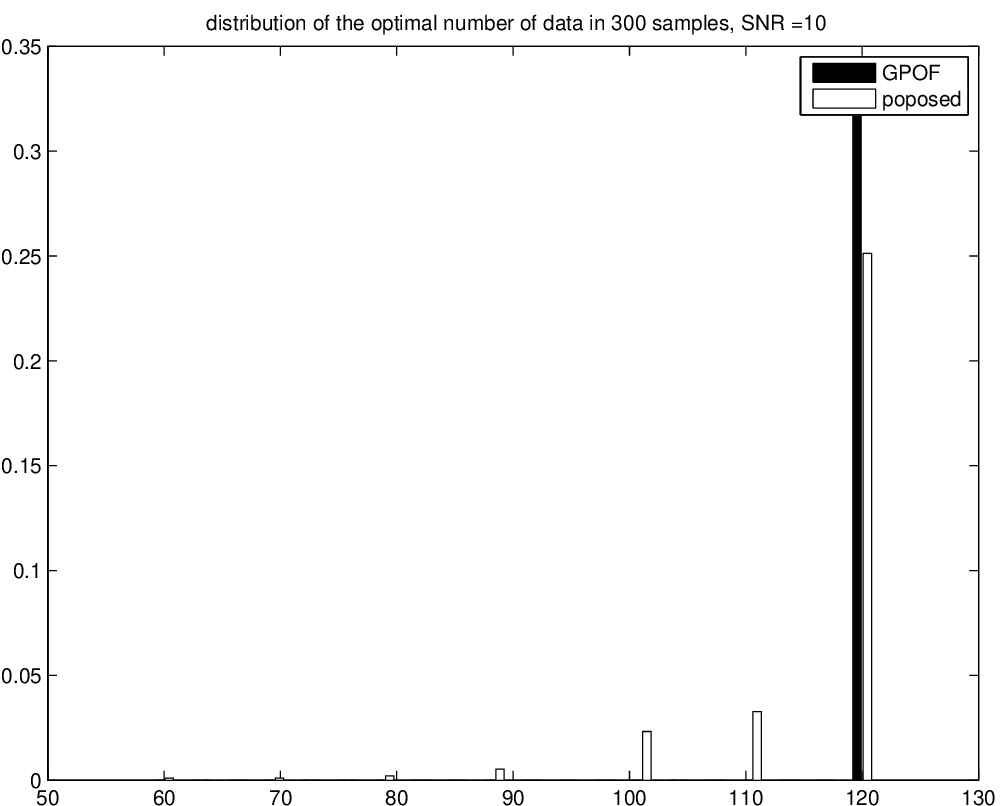,height=4.8cm,width=6cm}}
\hspace{.5cm}\fbox{\epsfig{file=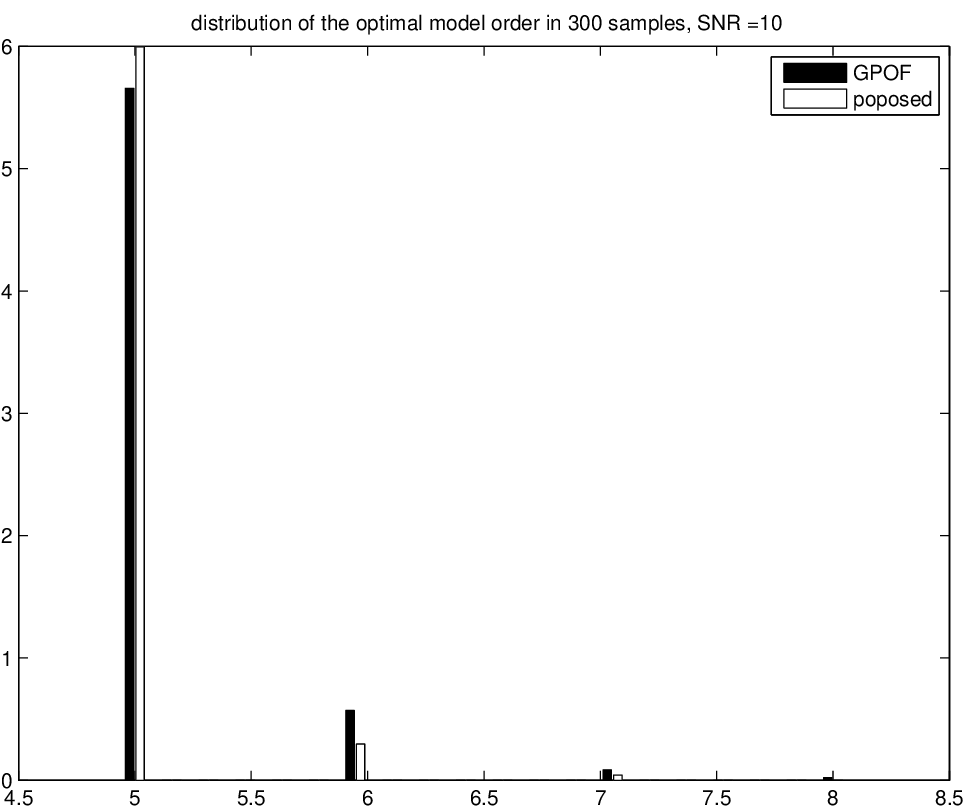,height=4.8cm,width=6cm}}}
\vspace{.2cm}
\hspace{1.8cm}\centerline{\fbox{\epsfig{file=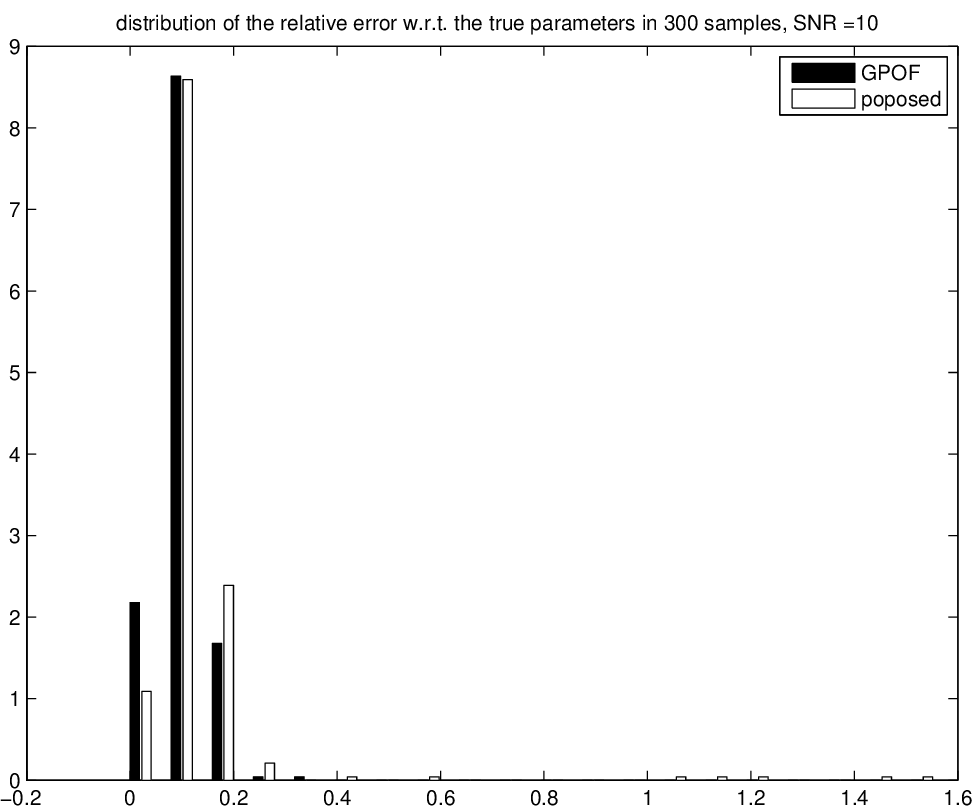,height=10cm,width=12.8cm}}
}
\end{center}
\caption{The empirical distributions over the $300$ replications of
$n_{ott}(\sigma,\cdot)$ (top left), $p_{ott}(\sigma,\cdot)$ (top right) and $E(\sigma,\cdot)$ (bottom) for $\sigma=\sqrt{2}/10$.   }  \label{lfig74}
\end{figure}

The average relative MSEs and the corresponding  cardinalities $N_\sigma$ are reported in the third and fourth rows of Table \ref{tb1}.
In fig. \ref{lfig71} the empirical distribution  of
$n_{ott}(\sigma,\cdot)$, $p_{ott}(\sigma,\cdot)$ and $E(\sigma,\cdot)$ were reported for  $\sigma=2\sqrt{2}$ for the standard and the proposed method. In figs. \ref{lfig72},\ref{lfig73},\ref{lfig74} the same was done for $\sigma=\sqrt{2},\sigma=\frac{\sqrt{2}}{3},\sigma=\frac{\sqrt{2}}{10}.$

As noted before if $n\le 100$ a super-resolution problem arises. Therefore we tried the standard and the proposed method with $n_{orig}=80$ and $n_0=30 $, $\sigma=2\sqrt{2},\sqrt{2},\frac{\sqrt{2}}{3},\frac{\sqrt{2}}{10}$ and $h=1,\dots,N$  and the hyperparameters  given in Table \ref{tb0}.
The results are reported in Table \ref{tb11}.

\begin{table}[tbh]
\begin{scriptsize}
\begin{center}
\begin{tabular}{|c|c|c|c|c|}
\hline\hline
&$SNR=0.5$&$SNR=1$&SNR=3&SNR=10\\
\hline
$MSE(standard)$&1.389&1.043&0.357&0.108\\
\hline
$N_\sigma$&98&149&263&300\\
\hline\hline
$MSE(proposed)$&0.905&0.707&0.369&0.155\\
\hline
$N_\sigma$&13&61&170&232\\
\hline
\end{tabular}
\end{center}
\caption{Super-resolution problem. Standard method: relative MSEs (first row) averaged over $N_\sigma$ (second row) replications. Proposed method: relative MSEs (third row) averaged over $N_\sigma$ (fourth row) replications. }\label{tb11}
\end{scriptsize}
\end{table}

From fig.\ref{lfig71},\ref{lfig72},\ref{lfig73},\ref{lfig74} and Table \ref{tb1} and  \ref{tb11} we conclude that results provided by the standard and the proposed method are similar for moderate or large SNRs (SNR$=3,10$). When the SNR is small (SNR$=0.5,1$) the proposed method is able to better identify the correct model order and hence, when this happens, better parameters estimates are obtained. Moreover in a few instances the proposed method can solve satisfactorily super-resolution problems ($MSE<1$) even for low SNRs.

Finally we used the proposed procedure for solving two problems discussed in \cite{dsp10} (see there for details) in order to appreciate the advantages of the new procedure w.r. to the original one. Among the problems afforded in \cite{dsp10} the most difficult ones are the interpolation and extrapolation of time series reported in \cite{cats} and the shape from moments problem.

\begin{figure}
\begin{center}
\hspace{1.7cm}{\fbox{\psfig{file=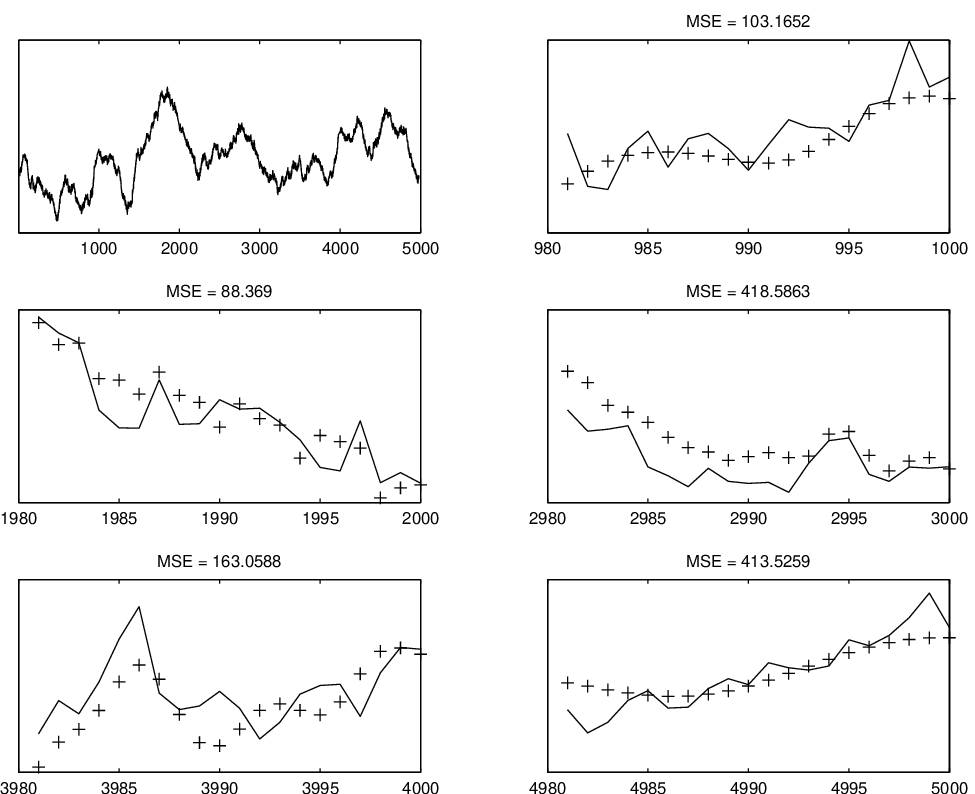,height=16cm,width=16cm}}}
\end{center}
\caption{Top
left: time series with five missing intervals. True values on each
interval (-); interpolated values (+). Total MSE on the first four
intervals = 193. Total MSE on the five intervals = 237. }
\label{fig9}
\end{figure}

The first problem copes with
a time series of $5000$ samples with $100$ missing
values at times $981-1000,
1981-2000,2981-3000,3981-4000,4981-5000$. Therefore we want
to solve four interpolation and one extrapolation problems.
As the data are synthetic the truth is known and the
results obtained by $17$ methods are reported in
\cite{cats} where the mean squared error (MSE) for the
interpolation problems and the interpolation +
extrapolation problems are reported. As in \cite{dsp10} we apply the
method to the residual obtained by subtracting a
smoothing cubic spline from the data. In fig.\ref{fig9} top left
the full time series with missing data is plotted. The
other plots show the true values and the reconstructed ones
on each missed data interval. The $MSE_{100}=237$ and
$MSE_{80}=193$ have to be compared with  $MSE_{100}=270$ and
$MSE_{80}=195$ obtained in \cite{dsp10} and with  $MSE_{100}=408$ and
$MSE_{80}=222$ which are the best results obtained in
\cite{cats} by two different methods among the $17$
considered. A slight improvement over the results reported in \cite{dsp10} can be noticed. However the most relevant fact is that these results were obtained by the black box procedure with the hyperparameters given in Table \ref{tb0}, the only data dependent information to provide are an estimate of the noise variance $\sigma^2$ and the initial number of data to use $n_0$.

\begin{figure}
\begin{center}
\hspace{1.7cm}{\fbox{\psfig{file=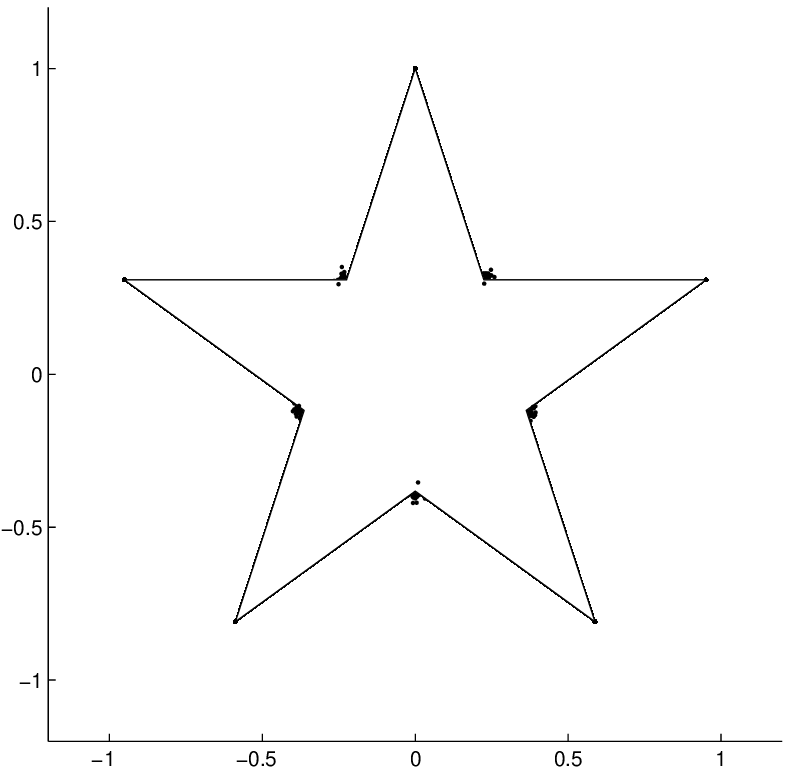,height=16cm,width=16cm}}}
\end{center}
\caption{Estimates of the vertices of the star shaped
polygon obtained by the proposed method on $N=100$ replications
with   $\sigma=1.e^{-4}$. }\label{lfig10}
\end{figure}

The second problem is the reconstruction of a non-degenerate polygon $\mathcal{P}$ from its complex moments. In  \cite{davis,gmv} there was shown that the $p$ vertices
$\xi_1,\dots,\xi_p$ of $\mathcal{P}$
 and its complex moments
$\mu_k,k=0,1,\dots,2p-1$ are related  by
$$k(k-1)\mu_k=k(k-1)\int_{\mathcal{P}}z^k dx\; dy=\sum_{j=1}^p c_j\xi^j,\;\;\mu_0=\mu_1=0$$
where
$$c_j=\frac{i}{2}\left(\frac{\overline{\xi}_{j-1}-\overline{\xi}_j}{\xi_{j-1}-\xi_j}-
\frac{\overline{\xi}_{j}-\overline{\xi}_{j+1}}{\xi_{j}-
\xi_{j+1}}\right)$$ assuming that the vertices are arranged
in counterclockwise direction in the order of increasing
index and extending the indexing of the $\xi_j$ cyclically
so that $\xi_0=\xi_p$, $\xi_1=\xi_{p+1}$. Therefore to
identify the polygon (i.e. its vertices) from its complex
moments is equivalent to solve a CEIP for the data
$s_k=k(k-1)\mu_k$. The proposed procedure was applied for
solving this  problem on a star shaped polygon for $\sigma=10^{-3},10^{-4},10^{-5}$ by a simulation experiment involving $N=100$ independent
replications and $n_{orig}=101$ noisy moments.
In Table \ref{tb4} the results obtained by the proposed method and those reported in
\cite[Table 2, first three lines]{dsp10} are given. For comparison also the results given in \cite{emg} in the far more easy case when the number of vertices is known are reported.  The root mean squared error
(RMSE) averaged over all parameters $\xi_j$ is computed by
$$RMSE=\frac{1}{p}\sum_{k=1}^p\sqrt{ \frac{1}{N}
\sum_{j=1}^{N}|\xi^{(j)}_k-\hat{\xi}^{(j)}_k |^2 }. $$
In fig.\ref{lfig10} the estimated $\xi_j$ for $\sigma=10^{-4}$ are plotted. We notice that the results obtained with the new procedure without knowing $p$ are much better of those reported in \cite{dsp10} in both cases when $p$ is known and unknown and in \cite{emg}  when $p$ is known. We stress that  in this experiment too the hyperparameters give in Table \ref{tb0} were used and $n_0=n_{orig}$ because the true signal is made of pure sinusoids.

\begin{table}[tbh]
\begin{scriptsize}
\begin{center}
\begin{tabular}{|c|c|c|c|c|c|}
\hline\hline
$\sigma$& RMSE & RMSE  \cite{dsp10}, $p$ unknown &RMSE  \cite{dsp10}, $p$ known &RMSE \cite{emg}, $p$ known\\
\hline\hline
1e-3&1.0e-2&1.07e-1&3.68e-2&5.74e-2\\
\hline
1e-4&3.2e-4&7.62e-2&1.02e-2&1.74e-2\\
\hline
1e-5&3.3e-5&2.98e-2&1.05e-3&1.71e-3\\
\hline \hline
\end{tabular}
\end{center}
\caption{For the star shaped polygon  the RMSE averaged over all the vertices obtained in the proposed procedure when $p$ is unknown, in \cite{dsp10} when $p$ is unknown, in \cite{dsp10} when $p$ is known,
in \cite{emg} when $p$ is known for $\sigma=1e^{-3},1e^{-4},1e^{-5}$ is reported. }
\label{tb4}
\end{scriptsize}
\end{table}

\section{Conclusions}
A classic approximation problem which is at the core of many ill posed inverse problems arising in many application fields is revisited and a new stochastic approach is considered to overcome the drawbacks of standard methods. It turns out that some  tools developed in the framework of the theory of random matrices, such as the condensed density of the generalized eigenvalues of a pencil of matrices, provides a deep insight on the structure of the approximation problem. Coupling this information with a stochastic perturbation approach, the bias which affects standard estimators based on Maximum Likelihood can be controlled and  a solution with better statistical properties, than those provided by standard methods, can be computed.  The proposed method  depends on two critical hyperparameters. One of them can be chosen in an optimal way  by partially heuristic considerations; the other one is chosen among a finite set of candidates by a simple selection procedure based on a measure of stationarity of the residuals. A few not critical hyperparameters must  be chosen too, which however turn out to be robust w.r. to the data and can be assigned once and for all independently of the application. These claims are checked by numerical experiments which improve over published results.

\newpage

\newpage

\end{document}